\documentclass[pra,twocolumn,showpacs,amsmath,amssymb]{revtex4}
\usepackage{graphicx}
\usepackage{amsfonts}
\usepackage{amsmath}
\usepackage{amssymb}
\usepackage{color}
\usepackage{bm}
\usepackage{bbm}
\usepackage{enumerate}

\graphicspath{{figures/}} 
\usepackage{amsfonts, amsmath, amsthm, amssymb} 
\usepackage{array}
\usepackage[colorlinks,bookmarks=false,citecolor=blue,linkcolor=red,urlcolor=blue]{hyperref}

\newcommand{\be}{\begin{equation}}
\newcommand{\bee}{\begin{equation*}}
\newcommand{\ee}{\end{equation}}
\newcommand{\eee}{\end{equation*}}
\newcommand{\bearre}{\begin{eqnarray*}}
\newcommand{\eearre}{\end{eqnarray*}}
\newcommand{\bearr}{\begin{eqnarray}}
\newcommand{\eearr}{\end{eqnarray}}

\newcommand\QMF{\stackrel{{\mbox{\rm \tiny QMF}}}{\approx}}

\newcommand\myeq{\stackrel{{\mbox{\rm \tiny $\beta \Delta E_1 \gg 1$}}}{\approx}}

\begin{document}

\title{Quantum mean-field approximation for lattice quantum models: truncating quantum correlations, and retaining classical ones}

\author{Daniele Malpetti$^1$
 and Tommaso Roscilde$^{1,2}$
 }

\affiliation{$^1$ Laboratoire de Physique, CNRS UMR 5672, Ecole Normale Sup\'erieure de Lyon, Universit\'e de Lyon, 46 All\'ee d'Italie, 
Lyon, F-69364, France}
\affiliation{$^2$ Institut Universitaire de France, 103 boulevard Saint-Michel, 75005 Paris, France}

\date{\today}

\begin{abstract}
The mean-field approximation is at the heart of our understanding of complex systems, despite its fundamental limitation of completely neglecting correlations between the elementary constituents.
In a recent work [D. Malpetti and T. Roscilde, arXiv:1605.04223] we have shown that in quantum many-body systems at finite temperature, two-point correlations can be formally separated into a thermal part, and a quantum part -- and that generically quantum correlations decay exponentially at finite temperature, with a characteristic, temperature-dependent quantum coherence length. The existence of these two different forms of correlation in quantum many-body systems suggests the possibility of formulating an approximation which affects quantum correlations only, without preventing the correct description of classical fluctuations at all length scales. Focusing on lattice boson and quantum Ising models, we make use of the path-integral formulation of quantum statistical mechanics to introduce such an approximation -- that we dub \emph{quantum mean-field} (QMF) approach, and which can be readily generalized to a cluster form (cluster QMF or cQMF). The cQMF approximation reduces to cluster mean-field theory at $T=0$,  while at any finite temperature it produces a family of systematically improved, semi-classical approximations to the quantum statistical mechanics of the lattice theory at hand. Contrary to standard MF approximations, the correct nature of thermal critical phenomena is captured by any cluster size. In the two exemplary cases of the two-dimensional quantum Ising model and of two-dimensional quantum rotors, we study systematically the convergence of the cQMF approximation towards the exact result, and show that the convergence is typically linear or sub-linear in the { boundary-to-bulk ratio} of the clusters as $T\to 0$, while it becomes faster than linear as $T$ grows. These results pave the way towards the development of semi-classical numerical approaches based on an approximate, { yet} systematically improved account of quantum correlations. 
\end{abstract}

\maketitle


\section{Introduction}

 The quantitative understanding of quantum many-body systems is one of the most challenging issues of modern theoretical physics, with a huge range of applications from materials science to high-energy physics. A generic method in the study of the quantum many-body problem -- and of many-body systems of all kinds -- is the mean-field (MF) approximation \cite{Goldenfeldbook}, which accounts for collective phenomena by considering only the correlations between the average behavior of the constituents, and discarding any form of correlation among fluctuations.
 Mean-field theory has many different declinations, depending on the physical system at hand. In this paper we shall focus on bosonic and spin models on a lattice; in this context the elementary degrees of freedom are identified with the bosonic modes or spin degrees of freedom attached to each individual lattice site \cite{Fisheretal1989,RokhsarK1991} (or cluster of sites, as in the context of cluster MF theory \cite{Bethe1935, Peierls1936}).  
 
 Decoupling the fluctuations of different lattice sites (or clusters thereof) opens the door to inexpensive, and often analytical, computation of ground-state and thermal phase diagrams of complex many-body systems. In numerous cases, this approach has the invaluable merit of capturing correctly their features at a semi-quantitative level. Nonetheless, as it is well known, the MF approximation, as well as its cluster extensions, completely miss all the physical aspects of many-body systems which are dominated by long-range fluctuations, such as the nature of critical points below the upper critical dimension, or extended critical phases such as the low-temperature Kosterlitz-Thouless superfluid phase of $U(1)$ symmetric systems in two dimensions. 
 
 In the case of classical many-body systems, the generic remedy to the shortcomings of MF approximation is the use of numerical simulations, such as the ones based on the Monte Carlo method, fully capturing the long-wavelength fluctuations dominating the behavior of the system at critical points \cite{NewmanBbook}. In the context of quantum many-body systems, such a remedy is unfortunately limited to models which do not suffer from a sign problem \cite{ChandrasekharanW1999,HeneliusS2000}, something which excludes a large variety of systems of great importance for the ongoing experiments in condensed matter or in atomic physics, such as frustrated quantum magnets and bosons in gauge fields. At the same time, one can argue that the long-wavelength fluctuations -- missed by the cluster mean-field theory and dominating the physics at finite temperature -- are primarily of thermal origin. Therefore a partial account of quantum effects (namely a semiclassical approach) should be generically able to faithfully describe the correlation properties and the critical phenomena at finite temperature. 
 
 In a recent work \cite{MalpettiR2016} we have shown that thermal and quantum correlations between subsystems $A$ and $B$ of an extended quantum system can be formally separated at equilibrium. The thermal part can be identified with the response function of $A$ to a field applied on $B$, while the quantum part can be identified with the difference between the correlation function and the response function (or ``fluctuation-dissipation discord"). 
 The main aim of this paper is to show that the path-integral formulation of quantum statistical mechanics allows to associate the thermal and quantum correlations with precise geometrical properties of imaginary-time paths.
 This insight allows one to formulate a (cluster) MF approximation \emph{restricted to quantum correlations} only, hereafter denoted (cluster) quantum mean-field (cQMF) approximation. The rationale of this approximation is rooted in the peculiar decay of quantum field correlations with distance, which is generically found to be much faster than that of thermal correlations: while thermal correlations govern the decay of the total correlations -- and can therefore exhibit a power-law decay or an exponential decay at finite temperature, depending on the phase of the system -- quantum correlations are found to exhibit generically a quantum coherence length $\xi_Q$ over which they decay exponentially \cite{MalpettiR2016}, and which is finite at \emph{any} finite temperature. 
 
 We give two concrete examples of the cQMF approximation by studying two paradigmatic quantum models whose thermodynamics can be solved via a path-integral approach (supplemented with a Monte Carlo simulation), and for which the cQMF can therefore be systematically implemented. The models of interest are the two-dimensional Ising model in a transverse field, featuring a finite-$T$ Ising transition which can be arbitrarily suppressed by the quantum effects induced by the field; and the two-dimensional quantum-rotor model { of interacting bosons}, featuring a Kosterlitz-Thouless transition which can also be arbitrarily suppressed by the interaction parameter.  
In both cases a standard MF approximation, neglecting correlations completely, greatly overestimates the critical temperature; it completely misses the correct critical behavior; and, even more seriously, in the case of quantum rotors it completely misses the extended critical nature of the low-temperature superfluid phase. On the contrary, the cQMF approximation describes correctly both the critical point and the low-temperature phase, as it only truncates the range of \emph{quantum} correlations to the size of the cluster, leaving intact the long-wavelength thermal fluctuations. When the cluster size is much larger than $\xi_Q$, the approximation can be very accurate. Our results offer a new angle of attack to strongly correlated spin and bosonic models, which can be naturally exploited by numerical approaches, as we shall further elaborate upon in a future publication \cite{AFMC}. { Other approximation schemes exist addressing uniquely the quantum fluctuations, such as the pure-quantum self-consistent harmonic approximation \cite{Cuccolietal1995} of the Feynman-Kleinert approach \cite{FeynmanK1986}. But, unlike other approaches, it accounts for fully non-linear quantum fluctuations at short-range, which, as we shall see, are essential to determine the quantum renormalization of classical behavior in models sufficiently fare from a quantum critical point.} 
  
  The structure of the paper is as follows: Sec.~\ref{sec:MF} introduces the cluster mean-field approach; Secs.~\ref{sec:PI} and \ref{sec:PI_Ising} and reformulate the mean-field approach within the context of coherent-state path-integrals for lattice-boson models and Ising-spin path integrals for quantum Ising spins; Sec.~\ref{sec:QMF} introduces the cQMF approximation; Sec.~\ref{s.cqcorr} defines the separation between classical/thermal and quantum correlations, laying the foundations of the cQMF approach; Secs.~\ref{s.Ising_QMF} and \ref{s.QR_QMF} finally apply the cQMF approach to quantum Ising spins and quantum rotors on the square lattice, respectively; conclusions are drawn in Sec.~\ref{s.conclusions}.

\section{Cluster mean-field approximation for ground and thermal states}
\label{sec:MF}

 Mean-field (MF) theory has the generic effect of decoupling the fluctuations of selected degrees of freedom in the system, and coupling the degrees of freedom only through their average value. For quantum many-body systems MF theory has many different declinations, depending on the decoupling scheme -- the most famous one possibly being Hartree-Fock theory for fermionic particles \cite{Vignalebook}, in which the mean-field decoupling occurs between extended modes of the fermionic field (such as electronic Bloch waves in a crystal or atomic orbitals in an atom). In this paper we shall rather focus on bosonic models on a lattice, encompassing lattice gases of bosonic particles, as well as spin models. In this context, the MF approximation is most successful when it decouples lattice sites, namely well localized modes of the Bose field. When dealing with the ground state of the system, the MF approximation is therefore equivalent to a variational Ansatz in the form of a factorized state
 \begin{equation}
 |\Psi_{\rm MF}\rangle = \otimes_c |\Psi_c\rangle
 \end{equation}
 where, in the standard form, $c$ is the index of lattice sites. In its most general formulation, $c$ can be the index of a cluster of sites, whose periodic repetition tiles the entire lattice; in this case one speaks of a cluster mean-field (cMF) approximation. The cMF approximation discards any form of entanglement among degrees of freedom belonging to different clusters -- with entanglement representing the most general form of correlation in pure quantum states.  This approximation is of course exact in a lattice with infinite connectivity, in which the entanglement (a ``monogamous" resource \cite{Horodecki2009}) is so spread between all degrees of freedom that it becomes negligible.  
 
  The cMF approximation has been recently applied to many lattice-boson and spin models \cite{McIntoshetal2012,Luehmann2013,Yamamotoetal2013, Trousseletetal2014, Yamamotoetal2015}, as it can provide a semi-quantitative account of many-body effects in models that are hardly tractable or intractable with numerically exact methods, such as two-dimensional models with frustration.  In particular an extrapolation in the size of the clusters (cluster-size scaling) makes it possible to guess the behavior of the system in the limit of infinite clusters (corresponding to the exact result). In particular, the extrapolation used in the literature is generically polynomial in the surface-to-bulk ratio $\lambda$, defined as 
  \begin{equation}
  \lambda  = \frac{N_{\rm ext}}{N_{\rm int} + N_{\rm ext}}
  \end{equation}
  where $N_{\rm ext}$ is the number of bonds linking the degrees of freedom within the cluster to those external to the cluster, while $N_{\rm int}$ is the number of bonds within the cluster. Hence $\lambda = 1$ for the basic (single-site) mean-field approximation, while $\lambda = 0$ corresponds to the exact result of a single cluster covering the whole system. 
 
  When considering finite temperatures, the standard extension of the cMF approximation implies the minimization of the free energy 
  {
  \begin{equation}
  F[\rho;T] = {\rm Tr}(\rho {\cal H} + k_B T \rho \log \rho)
  \label{e.F}
\end{equation}}
 (where ${\cal H}$ is the Hamiltonian, $T$ is the temperature and $k_B$ the Boltzmann constant) with a factorized density matrix $\rho$
 \begin{equation}
 \hat\rho_{\rm MF} = \otimes_c \hat\rho_c~.
 \label{e.rhoMF}
 \end{equation}
 Such an approximation lies at the heart of our understanding of critical phenomena \cite{Goldenfeldbook}, but it cannot reproduce quantitatively the thermodynamics, and, most importantly, the critical behavior of any system below the upper critical dimension. Indeed the Ansatz in Eq.~\eqref{e.rhoMF} neglects any form of correlation between different clusters, both for thermal fluctuations as well as for quantum ones.  
 
  The main scope of our paper is to show that the MF approximation can in fact be restricted to quantum fluctuations only, preserving the accurate description of long-wavelength thermal fluctuations. This allows to describe correctly finite-temperature critical phenomena -- governed by thermal fluctuations -- and to reproduce quantitatively the renormalization of the characteristic energy scales of the system caused by (short-ranged) quantum fluctuations.  
 
 \section{Different approximations for lattice bosons from the path-integral perspective}
\label{sec:PI}

In this section we shall review most common approximations to the thermodynamics of lattice bosons from the point of view of the coherent-state path-integral approach to the partition function, which will serve as a fundamental basis to introduce our QMF approximation in Sec.~\ref{sec:QMF}. 

 For concreteness, we shall focus our attention on a lattice spinless boson model
 \begin{eqnarray}
 \hat{\cal H}(\{ \hat b_i, \hat b_i^{\dagger}\}) & = &  \sum_{i\neq j} \left [ J_{ij} \hat b_i^{\dagger} \hat b_j    +  \frac{V_{ij}}{2} \hat n_i  \hat n_j \right ] \nonumber \\
 & +& \sum_i \left [ \frac{U}{2} \hat n_i( \hat n_i-1) - \mu \hat n_i \right]
 \label{e.H}
 \end{eqnarray}
 with a first term comprising off-site hopping ($J_{ij}$ - possibly complex) and interactions ($V_{ij}$), and an on-site interaction ($U$) and chemical potential ($\mu$) term. The indices $i$ and $j$ run over the sites of an arbitrary periodic lattice.  
 More general models (including spin degrees of freedom, non-abelian gauge fields, etc.) can also be treated similarly, but we refrain from considering them for the sake of simplicity.

 A coherent-state path-integral approach \cite{NegeleO1988} to the equilibrium behavior of the system expresses its partition function as 
 \begin{equation}
 {\cal Z} = \int {\cal D}[\{\phi_i(\tau)\}] e^{-S[\{\phi_i(\tau)\}]}
 \end{equation}
 where 
 \begin{equation}
 S[\{\phi_i(\tau)\}] =  \int_0^{\beta} d\tau ~\left [ \sum_i \phi_i^*(\tau) \frac{\partial}{\partial \tau} \phi_i(\tau) + {\cal H}(\{\phi_i(\tau),\phi^*_i(\tau)\}) \right]
 \label{e.S}
 \end{equation}
  is the Euclidean action for the complex field $\phi_i(\tau)$ dependent on imaginary time $\tau\in[0,\beta]$ where $\beta = (k_B T)^{-1}$. We would like to stress at this point the (rather obvious) fact that the coupling terms among different sites in Eq.~\eqref{e.S} are uniquely contained in ${\cal H}$, and they are therefore completely local in imaginary time: this observation shall be crucial in the following.   
  
  \subsection{Classical field theory}
  
  The quantum nature of the complex field $\phi_i(\tau)$ stems fundamentally from its imaginary-time dependence. In this respect, the classical approximation to the bosonic theory amounts to taking static fields 
  \begin{equation}
  \phi_i(\tau) \to \Psi_{i} ~~~ \text{(classical approximation)} ~.
  \end{equation} 
 At $T=0$ the classical-field (CF) approximation corresponds to Gross-Pitaevskii theory \cite{StringariPbook} -- equivalent to searching the ground state in the form of a product of coherent states
 \begin{equation}
 |\Psi_{\rm GP}\rangle = \otimes_i ~ e^{\Psi_i \hat b_i^{\dagger} - \Psi_i^* \hat b_i} ~|0\rangle 
 \end{equation}
  via the minimisation of the Gross-Pitaevskii (GP) functional  $\langle \Psi_{\rm GP} | \hat {\cal H} |  \Psi_{\rm GP}\rangle = {\cal H}(\{\Psi_i,\Psi_i^*\})$. Clearly this approximation can only reproduce condensate ground states, and fails completely in the case of bosonic insulators. At $T>0$ the classical approximation produces a complex classical field theory \cite{Castin2004} with Boltzmann weights $\exp[-\beta {\cal H}(\{\Psi_i,\Psi_i^*\})]$, whose partition function therefore reads 
 \begin{equation}
 {\cal Z} = \int \left ( \prod_j \frac{ d \Psi_j d \Psi_j^*}{2\pi i}  \right ) e^{-\beta {\cal H}(\{\Psi_i,\Psi_i^*\})}~.
 \label{eq:CF}
 \end{equation} 
 Such an approximation has been widely used in the context of weakly interacting Bose-Einstein condensates \cite{Blakieetal2008}.  The GP functional possesses the same symmetries as the original quantum Hamiltonian, and hence it can generally be expected to possess the same phase diagram as well, provided that the ground state is a condensate, and that quantum effects are indeed irrelevant in determining the critical behavior. Obviously any quantum effect renormalizing the value of the order parameter(s) and the position of the critical point(s) is neglected. 
  
 \begin{figure*}[ht!]
\mbox{
\includegraphics[width = 0.5\linewidth]{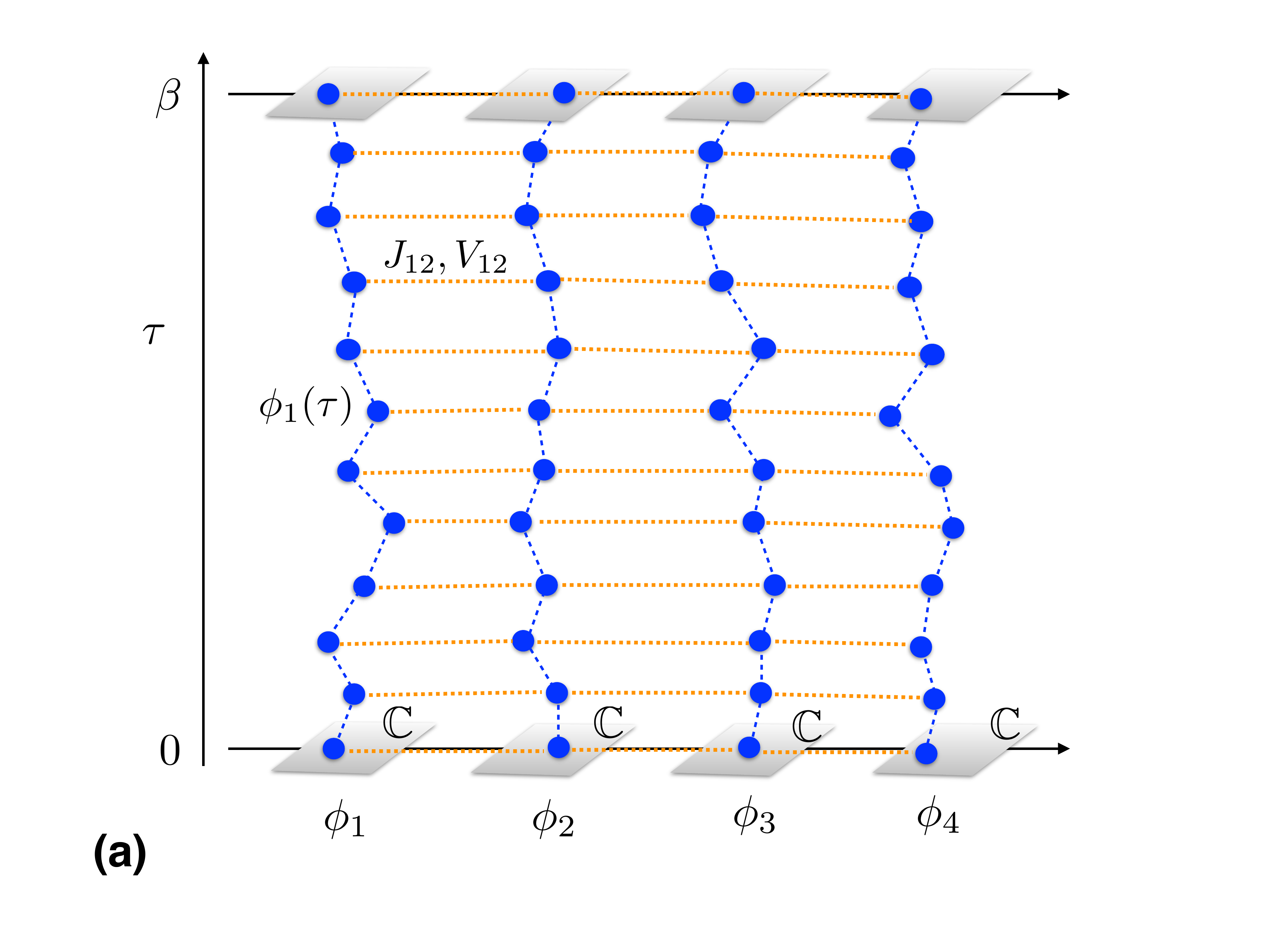}
\includegraphics[width = 0.5\linewidth]{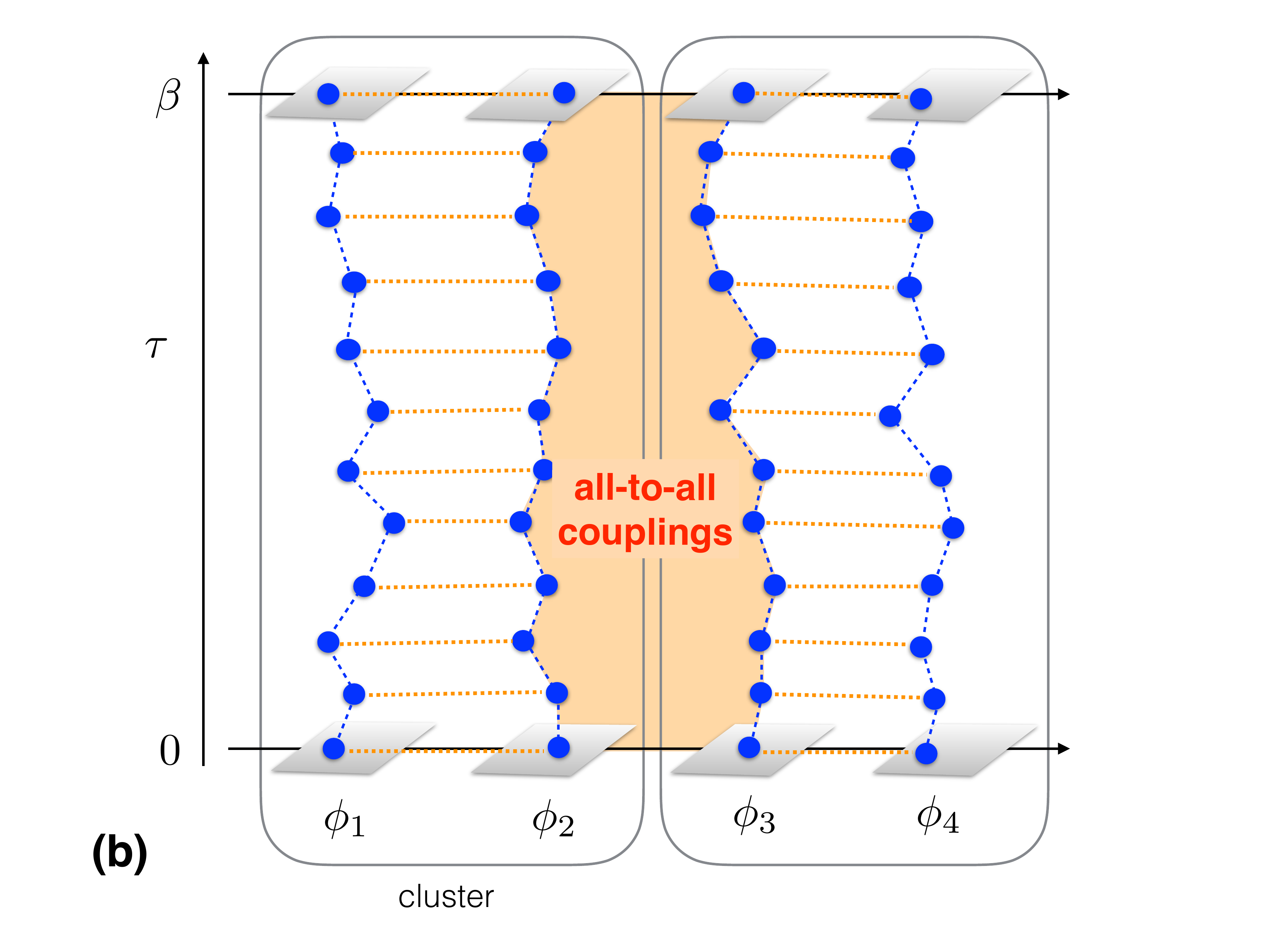}}
\caption{Cartoon of the cluster quantum mean-field (cQMF) approximation: (a) Coherent-state path integral configuration: to each site one associates a variable $\phi_i(\tau)$ in the complex plane $\mathbb{C}$, evolving in imaginary time $\tau$, and coupled to the other variables \emph{instantaneously} in imaginary time via the hopping $J_{ij}$ and the off-site interactions $V_{ij}$ (dashed orange lines);  (b) the cQMF approximation consists in dividing the system into clusters (rounded boxes) and approximating the instantaneous $J_{ij}$ and $V_{ij}$ couplings between clusters with all-to-all couplings in imaginary time, namely couplings between the imaginary-time-averaged variables $\bar\phi_i$ and ${\bar n}_i$ (see text) -- but fully preserving the spatial structure of the couplings in real space.}
\label{f.QMF}
\end{figure*}

  \subsection{Cluster mean-field theory} 
  
 The cluster mean-field theory approach generalizing the Ansatz Eq.~\eqref{e.rhoMF} amounts to breaking the partition function into a product of cluster partition functions determined self-consistently, namely 
 \begin{equation}
 {\cal Z}_{\rm cMF} = \prod_c {\cal Z}_c
 \label{e.ZcMF}
 \end{equation}
 where 
\begin{equation}
 {\cal Z}_c =  \int {\cal D}[\{\phi_{i\in c}(\tau)\}] ~e^{-S^{\rm (MF)}_c}
 \end{equation}
 and 
 \begin{eqnarray}
 S^{\rm (MF)}_c[\{\phi_{i\in c}(\tau)\}]  =   \int d\tau ~{\cal H}_c(\{{\phi_{i\in c}(\tau),\phi^*_{i\in c}(\tau)}\}) && \nonumber  \\
 +   \int d\tau \sum_{i\in c, j\notin c} \left[-\left( J_{ij}  \phi_i^{*}(\tau) \langle \hat b_j \rangle + {\rm c.c.} \right) + V_{ij} n_i(\tau)  \langle \hat n_j \rangle \right ] && \nonumber \\
  -  \frac{\beta}{2} \sum_{i\in c, j\notin c} \left[-\left( J_{ij}  \langle \hat b^{\dagger}_i \rangle \langle \hat b_j \rangle + {\rm c.c.} \right) + V_{ij}   \langle \hat n_i \rangle \langle \hat n_j \rangle \right ] &&~~~
 \end{eqnarray}
 where ${\cal H}_c$ indicates the Hamiltonian of the form Eq.~\eqref{e.H} restricted to sites $i$ and $j$ belonging to the cluster $c$. Here $\langle ... \rangle$ indicates the thermal average, to be calculated self-consistently using the partition function in the form of Eq.~\eqref{e.ZcMF}. Moreover we have introduced the notation $ n_i(\tau) =: |\phi_i(\tau)|^2$. 
 
  As already mentioned in Sec.~\ref{sec:MF}, the cMF approximation amounts to positing that the density matrix of the system has a separable form, and to minimize the free energy with such a form. From the path-integral perspective, the cMF approach (unlike classical field theory) implies to retain the quantum nature of the bosonic field, as well as its correlations within each cluster, but to discard correlations of all forms between clusters. Hence the approximation trades the treatment of the field as a quantum degree of freedom for the absence of long-wavelength fluctuations, preventing the correct description of critical phenomena below the upper critical dimension.

  \section{Different approximations for quantum Ising spins}  
  \label{sec:PI_Ising}
   
  In this section we shortly revisit the approximation schemes described in the previous section for the case of Ising spins in a transverse field, with Hamiltonian
  \begin{equation}
  \hat{\cal H} = -\sum_{i<j} J_{ij} \hat\sigma_i^z \hat\sigma_j^z - \Gamma \sum_i \hat\sigma_i^x~.
  \end{equation}
   where $\hat\sigma_i^{\alpha}$ ($\alpha=x,y,z$) are Pauli matrices. 
   
   Making use of a Trotter-Suzuki decomposition of the density operator \cite{Suzuki1976} the partition function of the system can be conveniently mapped onto that of a classical Ising model in $(d+1)$ dimensions
  \begin{equation}
  {\cal Z} = \lim_{M\to\infty} \sum_{\{\sigma_{i,k}\}}e^{-S_I[\{\sigma_{i,k}\}]}
  \label{e.ZPIsing}
  \end{equation} 
  with  
  \begin{equation}
   S_I[\{\sigma_{i,k}\}] = - \sum_{k=1}^{M} \sum_{i<j} K_{ij} \sigma_{i,k} \sigma_{j,k} -  K_{\tau} \sum_{k=1}^{M} \sum_i \sigma_{i,k} \sigma_{i,k+1}
   \label{e.SI}
   \end{equation}
   where 
  $K_{ij} = \beta J_{ij}/M$ and $K_{\tau} = \frac{1}{2} |\ln \tanh(\beta \Gamma/M)|$. Here $\sigma_{i,k} = \pm 1$ are classical Ising variables, and $k$ is the index of discrete imaginary-time slices defining the extra dimension. 
  
  Obviously the approximation $\sigma_{i,k} =: \sigma_i$ (independent of $k$) corresponds to a classical approximation, in which the transverse field is neglected altogether. On the other hand, the cMF approximation amounts to taking for the partition function the cluster-factorized form of Eq.~\eqref{e.ZcMF} with 
  \begin{equation}
  {\cal Z}_c = \sum_{\{\sigma_{i\in c,k}\}} e^{-S^{\rm (MF)}_c[\{\sigma_{i\in c,k}\}]}
  \end{equation}
   { Within the cMF approach, the action takes the following form}
  \begin{eqnarray}
  S^{\rm (MF)}_c[\{\sigma_{i\in c,k}\}] & = & S_I[\{\sigma_{i\in c,k}\}] - \sum_k \sum_{i\in c, j\notin c} K_{ij} \sigma_{i,k} \langle \sigma_{j,k} \rangle \nonumber \\
  & + & \frac{1}{2} \sum_k \sum_{i\in c, j\notin c} K_{ij} \langle \sigma_{i,k} \rangle  \langle \sigma_{j,k} \rangle~.
  \end{eqnarray}
  Here again $\langle ... \rangle$ denotes the thermal average taken with the same partition function, implying a self-consistent treatment. It is important to remark that the cMF approximation only affects the space-like couplings $K_{ij}$. 
  
 \section{Quantum mean-field approximation}
\label{sec:QMF}

In the previous section we have seen that CF theory and cMF theory have very complementary purposes. The former describes correctly thermal fluctuations at all wavelengths, but within a fully classical approximation to the Bose field or spin field; while the latter partially accounts for quantum effects, but discards all fluctuations (quantum and thermal) for wavelengths longer than the cluster size. Evidently an approximation which can accomplish both goals -- accounting for thermal and quantum fluctuations appropriately -- would be highly desirable. The cluster quantum mean-field approximation (cQMF) is precisely designed towards this goal.  

  Classical correlations are easily incorporated in the cMF Ansatz for the density matrix, Eq.~\eqref{e.rhoMF}, by promoting the elementary separable form of the density matrix to the most general, separable form between several clusters \cite{Werner1989}, namely
 \begin{equation}
 \rho_{\rm sep} = \sum_{\{\Psi\}} ~p(\{\Psi\}) ~\left [ \otimes_c ~\hat\rho_c(\{\Psi\}) \right ]
\label{e.rhosep}
\end{equation}
where $\{\Psi\}$ is a set of (possibly continuous) variables parametrizing the form of the cluster density matrix $\hat\rho_c(\{\Psi\})$, and $p(\{\Psi\})\geq 0$ is the probability of the associated factorized form. The temperature is contained in the functional form of both $\rho_c$ and $p$. 
The variational minimization of the free energy Eq.~\eqref{e.F} with respect to the separable Ansatz of Eq.~\eqref{e.rhosep} is obviously arduous, given the vast range of different parameterizations $\hat\rho_c(\{\Psi\})$ and $p(\{\Psi\})$ that one could choose. Hence, instead of a variational approach, a physically motivated Ansatz appears as a most viable route. In order to do so, the path-integral representation of the partition function is again of crucial help.

\subsection{QMF within the coherent-state path-integral approach}

The quantum mean-field approximation (QMF) is formulated at the level of the path-integral formulation, and it exactly provides a classically correlated Ansatz { (in the sense of Ref.~\cite{Werner1989})} for the density matrix of the system of clusters. 
The hopping and off-site potential terms, coupling sites together in the action Eq.~\eqref{e.S}, are fully \emph{local} in imaginary time: the QMF approximation consists then in treating them as completely \emph{non-local}, namely imagining that full connectivity exists along the imaginary-time dimension. Calling $S^{\rm (hop)}_{ij}$ and $S^{\rm (pot)}_{ij}$ the terms of the action coupling sites $i$ and $j$ (with $i\neq j$), the QMF approximation amounts to taking
\begin{eqnarray}
& S^{\rm (hop)}_{ij}&[\phi_i(\tau),\phi_j(\tau)] = - \int_0^{\beta} d\tau ~  \left ( \phi_i^*(\tau) ~ J_{ij}~ \phi_j(\tau) + {\rm c.c.} \right)   \nonumber \\
 & \QMF & - \frac{1}{\beta} \int_0^{\beta} d\tau \int_0^{\beta} d\tau' ~  \left ( \phi_i^*(\tau) ~ J_{ij} ~ \phi_j(\tau') + {\rm c.c.} \right) \nonumber \\
 &= & - \beta ~\bar{\phi_i}^* ~J_{ij} ~\bar{\phi_j}     =  S^{\rm (hop)}_{ij}[\bar \phi_i,\bar \phi_j]
\label{e.QMFhop}
\end{eqnarray}

\begin{eqnarray}
& S^{\rm (pot)}_{ij}&[n_i(\tau),n_j(\tau)]   =  - \int_0^{\beta} d\tau ~  n_i(\tau) ~ V_{ij}~ n_j(\tau)  \nonumber \\
 & \QMF & - \frac{1}{\beta} \int_0^{\beta} d\tau \int_0^{\beta} d\tau' ~ n_i(\tau) ~ V_{ij} ~ n_j(\tau') \nonumber \\
 &= & - \beta  ~\bar{n}_i ~V_{ij} ~\bar{n}_j   = S^{\rm (pot)}_{ij}[\bar n_i,\bar n_j]
\label{e.QMFpot}
\end{eqnarray}
where we have introduced the time-averaged field and density: 
\begin{equation}
\bar{\phi}_i = \frac{1}{\beta} \int_0^{\beta} d\tau ~\phi_i(\tau) ~~~~~~~~ \bar{n}_i = \frac{1}{\beta} \int_0^{\beta} d\tau ~n_i(\tau)~.
\end{equation}
In other words, the QMF approximation amounts to substituting the coupling between the imaginary-time evolutions of the field $\phi_i$, and of its squared amplitude $|\phi_i|^2$, at different sites with the coupling between the \emph{averages} over such evolutions -- whence the concept of quantum (or imaginary-time) mean field. This approximation is also equivalent to replacing the instantaneous couplings in imaginary time with all-to-all couplings (see Fig.~\ref{f.QMF} for a cartoon).   

The cluster QMF (cQMF) approximation amounts then to divide the system into clusters, divide the $ij$ bonds into intra-cluster and inter-cluster ones, and apply the mean-field approximation in imaginary time to all inter-cluster couplings.
Under this approximation, the coherent state action takes the form
\begin{eqnarray}
 S[\{\phi_i(\tau)\}]  &\QMF&  \sum_c S_c[\{\phi_{i\in c}(\tau)\}]  \nonumber \\
 &+& \sum_{c < c'} ~\sum_{i\in c, j\in c'} S_{ij}[\bar{\phi_i},\bar{\phi_j};\bar{n}_i,\bar{n}_j]~ 
\end{eqnarray}
where we have introduced the cluster action
\begin{eqnarray}
S_c[\{\phi_{i\in c}(\tau)\}]  &=& \sum_{i\in c} S_i[\{\phi_i(\tau)\}] \\
&+& \sum_{i,j\in c} S_{ij}[\phi_i(\tau),\phi_j(\tau);n_i(\tau), n_j(\tau)] ~.\nonumber 
\end{eqnarray}
Here $S_i$ is the part of the action containing exclusively the on-site terms (on-site interaction and chemical potential), while $S_{ij} = S^{\rm (hop)}_{ij} + S^{\rm (pot)}_{ij}$ is the off-site part, which becomes
\begin{equation}
S_{ij}[\bar{\phi_i},\bar{\phi_j};\bar{n}_i,\bar{n}_j] = S^{\rm (hop)}_{ij}[\bar \phi_i,\bar \phi_j] + S^{\rm (pot)}_{ij}[\bar n_i,\bar n_j]
\end{equation}
under the QMF approximation. 

The path integral over coherent states $\int {\cal D}[\{ \phi_i(\tau)\}]$  can then be decomposed into two integrals: an integral over average values of the fields $\bar{\phi_i}$ and of the densities $\bar{n}_i$, and a path-integral over paths realizing those average values \cite{Cuccolietal1995,FeynmanHibbs}:
 \begin{eqnarray}
 {\cal Z}   \approx ~~~~~~~~~~~~~~~~~~~~~~~~~~~~~~~~~~~~~~~~~~~~~~~~~~~~~~~~~~~~~~\\
 \int  \left( \prod_j  \frac{d\bar \phi_j ~d\bar \phi^*_j~d\bar{n}_j}{2\pi i} \right )  e^{-\sum_{c\neq c'}\sum_{i\in c,j\in c'} S_{ij}[\bar{\phi_i},\bar{\phi_j};\bar n_i,\bar n_j]} \nonumber \\
\prod_c  \int_{\{\bar\phi_{i\in c}, \bar n_{i\in c}\}}  {\cal D}[\{ \phi_{i \in c}(\tau)\}] ~e^{-S_c[\{ \phi_{i\in c}(\tau) \}] }~. \nonumber ~~~~~~~~~~~~~
 \label{e.QMF}
 \end{eqnarray}
  
\subsection{cQMF for quantum Ising spins}  
  
  The cQMF approximation can be similarly formulated for quantum Ising spins based on the path-integral representation of the partition function of Eq.~\eqref{e.ZPIsing}. The cQMF approximation amounts to defining the time-averaged Ising spin
  \begin{equation}
  \bar \sigma_i = \frac{1}{M} \sum_{k=1}^M \sigma_{i,k}
  \end{equation}
  and replacing the space-like coupling between the $\sigma_{i,k}$ spins, which is local in imaginary time, with a coupling between averaged spins on bonds connecting two different clusters. The action takes therefore the form
  \begin{equation}
 S_{\rm I}[\sigma_{i,k}] \QMF \sum_c S_c[\{\sigma_{{i\in c},k}\}] + \sum_{c< c'}  S_{cc'}[\{\bar\sigma_{i\in c}\};\{\bar\sigma_{j\in c'}\}] 
  \end{equation}
  where $S_c = S_I[\{\sigma_{{i\in c},k}\}]$ is the action restricted to sites and links belonging to a single cluster, whereas
  \begin{eqnarray}
  S_{cc'}[\{\bar\sigma_{i\in c}\};\{\bar\sigma_{j\in c'}\}] &=& - \sum_{i\in c, j\in c'} J_{ij} \bar \sigma_i \bar \sigma_j \nonumber \\
  &=& - \sum_{i\in c, j\in c'} \frac{J_{ij}}{M^2} \sum_{k,k'} \sigma_{i,k} \sigma_{j,k'}~.
 \end{eqnarray}
 The resulting partition function takes then the form of an integral over average spins $\bar\sigma_i$, and a path integral with fixed averages 
 \begin{eqnarray}
 {\cal Z} &\approx& \sum_{\{\bar \sigma_i\}} e^{-\sum_{c\neq c'}  S_{cc'}[\{\bar\sigma_{i\in c}\};\{\bar\sigma_{j\in c'}\}]} \\
 &\prod_c& \sum_{\{\sigma_{i\in c},k\} | \{ \bar\sigma_{i\in c} \}} e^{-S_c[\{\sigma_{{i\in c},k}\}]}
 \label{e.cQMFIsing}
 \end{eqnarray} 
 where the symbol  $\{\sigma_{i\in c,k}\} | \{ \bar\sigma_{i\in c} \}$ indicates the set of $\sigma_{i,k}$ spin configurations for a cluster $c$, conditioned on the time average being $\bar\sigma_i$.
  
\subsection{cQMF approximation and classically correlated states}  
  
The  cQMF expression for the partition function can be immediately cast into the form of the trace of a cluster-separable density matrix, as in Eq.~\eqref{e.rhosep}. 

\subsubsection{Lattice bosons}
In the case of lattice bosons, the cluster-separable form for the density matrix is achieved by treating the inter-cluster action via a Hubbard-Stratonovich (HS) decomposition, introducing two auxiliary fields, a complex-valued one ($\Psi_i$) and a real-valued one ($\varrho_i$): 
\begin{widetext}
 \begin{eqnarray}
 e^{-\sum'_{ij} S^{\rm(hop)}_{ij}[\bar{\phi}_i,\bar{\phi}_j]}  & = & 
 \frac{1}{\det J'} \int \left (\prod_i \frac{\beta^2 d\Psi_i d\Psi_i^*}{2\pi i} \right) e^{-\beta \sum'_{ij} \Psi_i^{*} [(J')^{-1}]_{ij} \Psi_j - \beta \sum'_i(\Psi_i \bar\phi^*_i + \Psi^*_i \bar\phi_i)}  \\
  e^{-\sum''_{ij} S^{\rm(pot)}_{ij}[\bar{n}_i,\bar{n}_j]} & = &  \frac{1}{\det V''} \int \left (\prod_i \beta~ d\varrho_i \right ) e^{-\beta \sum''_{ij} \varrho_i [(V'')^{-1}]_{ij} \varrho_j - \beta \sum''_i \varrho_i {\bar n}_i}  ~.
 \label{e.HS}
 \end{eqnarray}
 \end{widetext}
Here the primed (and double-primed) sums $\sum'$ ($\sum''$) are restricted to those sites which are indeed involved in an inter-cluster bond for the kinetic energy (the potential energy), and the matrices $J'$ and $V''$ are connectivity matrices restricted to inter-cluster bonds (for the kinetic energy and potential energy respectively). We shall here assume that $J'$ and $V''$ are positive definite, so that the HS transformation is well defined  -- otherwise they can be appropriately redefined, leaving the essence of the present argument intact. 
 
 The HS transformation allows therefore to cast the partition function within the cQMF approximation into the separable form, ${\cal Z} \approx {\rm Tr} (\rho_{\rm cQMF})$ with (up to multiplicative constants)
\begin{widetext}
 \begin{equation}
 \rho_{\rm cQMF} \sim \int \left (\prod_j \frac{d\Psi_j d\Psi_j^* d\varrho_i}{2\pi i} \right) e^{- S_{\rm aux}[\{\Psi_i, \Psi_i^*;\varrho_i\}] } \left (~ \otimes_c \hat\rho_c[\{\Psi_{i\in c}, \Psi_{i \in c}^*;\varrho_{i\in c}\}] ~ \right )
  \label{e.rhoQMF}
 \end{equation}
 \end{widetext}
   where we have introduced the auxiliary-field action
   \begin{eqnarray}
   S_{\rm aux}[\{\Psi_i, \Psi_i^*;\varrho_i\}] &=& \beta {\sum}'_{ij} ~\Psi_i^{*} [(J')^{-1}]_{ij} \Psi_j \nonumber \\
   &+& \beta {\sum}''_{ij} ~\varrho_i [(V'')^{-1}]_{ij} \varrho_j
   \end{eqnarray}
  and the single-cluster density matrix
   \begin{equation}
   \hat\rho_c[\{\Psi_{i\in c}, \Psi_{i \in c}^*;\varrho_{i\in c}\}]  = e^{-\beta \hat{\cal H}_c^{\rm (eff)} [\{\Psi_{i\in c}, \Psi_{i \in c}^*;\varrho_{i\in c}\}]}
   \end{equation}
   with the effective single-cluster Hamiltonian
   \begin{equation}
   \hat{\cal H}_c^{\rm (eff)} = \hat{\cal H}_c - {\sum_{i \in c}}' \left( \Psi_i \hat{b}^*_i + \Psi^*_i \hat{b}_i \right) - {\sum_{i \in c}}'' \varrho_i \hat n_i  
   \end{equation}
   where $\hat{\cal H}_c$ is the physical Hamiltonian, Eq.~\eqref{e.H}, restricted to intra-cluster bonds (for the off-site terms) and cluster sites (for the on-site terms). Therefore the effective cluster Hamiltonian has the form of the physical Hamiltonian plus ``boundary" source terms (containing the auxiliary fields $\Psi_i, \Psi_i^*$) and a ``boundary" potential term (containing the auxiliary field $\varrho_i$) involving the sites coupled to other sites outside the cluster.   
  
\subsubsection{Quantum Ising spins}
     
   A HS transformation analog to that of Eq.~\eqref{e.HS}  
   \begin{eqnarray}
   e^{\sum'_{ij} {\beta J'}_{ij} \bar\sigma_i\bar\sigma_j} = ~~~~~~~~~~~~~~~~~~~~~~~~~~~~~~~~~~~~~~~~~~~~ \\
   \frac{1}{\det J'} \int \left ( \prod_i \beta d\xi_i \right)~ e^{-\beta \sum'_{ij} \xi_i [(J')^{-1}]_{ij} \xi_j} ~e^{-\beta \sum'_i \xi_i \bar\sigma_i } \nonumber
   \end{eqnarray}  
    brings the density matrix to the cluster-separable form  
  \begin{equation}
 \hat\rho_{\rm cQMF} \sim \int \left ( \prod_i d\xi_i \right )~ e^{-\beta \sum'_{ij} \xi_i [(J')^{-1}]_{ij} \xi_j}   ~\otimes_c \hat\rho_c[\{\xi_{i\in c}\}] 
  \label{e.rhoQMF_2}
  \end{equation}
  where the single-cluster density matrix reads
  \begin{equation}
  \hat\rho_c = e^{-\beta \hat{\cal H}^{\rm (eff)}_c[\{\xi_{i\in c}\}]}
  \end{equation}
  with the cluster effective Hamiltonian given by
  \begin{equation}
  \hat{\cal H}^{\rm (eff)}_c[\{\xi_{i\in c}\}] = -\sum_{i,j\in c} J_{ij} \hat\sigma_i^z \hat\sigma_j^z - \sum_{i\in c} \Gamma \hat\sigma_i - {\sum_{i \in c}}' \xi_i \hat\sigma^z_i
  \end{equation}   
  corresponding to the physical Hamiltonian augmented with a longitudinal ``boundary" field term involving the spins coupled to other spins outside the cluster. 
     
  \section{Thermal vs. quantum correlations}
  \label{s.cqcorr}
  
  We have seen in the previous section that the cQMF approximation allows to cast the density matrix of the system in the form of an operator describing intra-cluster couplings in a fully quantum-mechanical way, and accounting for classical correlations among such clusters. Hence, according to the definition of entanglement and non-separability of mixed states \cite{Werner1989, Horodecki2009}, the fundamental missing ingredient of the cQMF approximation is the \emph{entanglement between clusters}. In other words, the cQMF truncates quantum correlations to distances not exceeding the linear size of the clusters. The path-integral formulation of the partition function provides a novel insight into the meaning of entanglement between clusters. The neglect of correlations in imaginary time between clusters leads to their separability in the density matrix. Conversely, the existence of correlations of imaginary-time fluctuations between different degrees of freedom allows to exclude their separability in the form of Eq.~\eqref{e.rhoQMF} -- named \emph{Hamiltonian separability} in Ref.~\cite{MalpettiR2016}. The latter form of separability has an immediate physical significance: the density matrix is cluster-separable since clusters are governed by local Hamiltonians which are coupled via classical fields, $\Psi_i$ and $\varrho_i$ (for lattice bosons), and $\xi_i$ (for Ising spins). Disproving Hamiltonian separability does not imply disproving the most general form of separability in Eq.~\eqref{e.rhosep} -- which would be equivalent to proving entanglement, and which remains a challenging task. Nonetheless the absence of Hamiltonian separability is tightly related to the existence of quantum correlations, as pointed out in Ref.~\cite{MalpettiR2016}, and as we shall further elucidate here.

 The most important form of correlation in a bosonic field theory is the first-order correlation function $g(l,m) = \langle \hat b_l^\dagger \hat b_m \rangle$, which acquires the path-integral expression
  \begin{equation}
  g(l,m) = \frac{1}{\cal Z} \int ~{\cal D}[\{{\phi_i}(\tau)\}] ~\phi_l^*(\tau') \phi_m(\tau') ~e^{-S}
  \end{equation}
  for any time $\tau' \in [0,\beta]$. Clearly, the $g$ function probes field correlations which are instantaneous in imaginary time. It is then very natural to define \emph{thermal field correlations} as the correlations of the time-averaged fields
  \begin{eqnarray}
  g_T(l,m) &=& \frac{1}{\beta} \int d\tau ~\langle \hat b_l^{\dagger}(\tau) \hat b_m (0)\rangle \nonumber \\
  &=&  \frac{1}{\cal Z} \int ~{\cal D}[\{{\phi_i}(\tau)\}] ~\bar{\phi}_l^* \bar{\phi}_m ~e^{-S}~.
  \label{e.classcorr}
  \end{eqnarray} 
  Indeed we have seen before that two sites (or clusters) which are decoupled via the QMF approximation are correlated uniquely via their time-average fields, and they are separable, namely they are only classically correlated (in the sense of Ref.~\cite{Werner1989}).  
  
  As a consequence, it is then rather natural to define \emph{quantum field correlations}  \cite{MalpettiR2016} as the difference between total and thermal correlations 
  \begin{eqnarray}
 && g_Q(l,m) =  g(l,m) - g_T(l,m) \\
   &=&  \int {\cal D}\phi \left [ \frac{1}{\beta}\int d\tau' (\phi_l^*(\tau')-\bar{\phi}_l^*)(\phi_m(\tau')-\bar{\phi}_m) \right ]~\frac{e^{-S}}{\cal Z} \nonumber 
     \label{e.GQ}
  \end{eqnarray}
  (where ${\cal D}\phi = {\cal D}[\{{\phi_i}(\tau)\}]$)
namely quantum field correlations are correlations of the \emph{fluctuations of the quantum fields around their imaginary-time average}.

The above considerations obviously extend to any correlation function - namely, to correlation functions of order higher than one. For instance, when considering density-density correlations  
 $G(l,m) = \langle \delta \hat n_l \delta \hat n_m \rangle$ with $\delta \hat n = \hat n - \langle n \rangle$, the quantum-correlation part takes the form    
  \begin{eqnarray}
 && G_Q(l,m)  =   G(l,m) - \frac{1}{\beta} \int d\tau \langle \hat n_l(\tau) \hat n_m(0) \rangle  \\
  &=& \int {\cal D}[\{{\phi_i}(\tau)\}] \left[ \frac{1}{\beta} \int d\tau' (n_l(\tau')-\bar{n}_l)(n_m(\tau') -\bar{n}_m) \right ] ~\frac{e^{-S}}{\cal Z} \nonumber
  \end{eqnarray}
 
 In the case of Ising spins, the most important correlation function is $C(i,j) =  \langle \hat\sigma_i^z \hat\sigma_j^z \rangle - \langle \hat\sigma_i^z \rangle \langle \hat\sigma_j^z \rangle$, with the associated quantum correlation function 
\begin{eqnarray}
 && C_Q(l,m) = C(l,m) - C_T(l,m)  \\
 &=& \langle \hat\sigma_l^z \hat\sigma_m^z \rangle - k_B T \int d\tau \langle \hat\sigma_l^z(\tau) \hat\sigma_m^z(0) \rangle \nonumber \\
&=& \lim_{M\to\infty} \sum_{\{\sigma_{i,k}\}} \left [\frac{1}{M} \sum_k  (\sigma_{l,k}-\bar\sigma_l) (\sigma_{m,k}-\bar\sigma_m) \right] \frac{e^{-S_I}}{\cal Z} \nonumber~.
 \end{eqnarray}

To further convince oneself of the validity of the above distinction between thermal and quantum field correlations, it suffices to look back at the QMF approximation once more: within this approximation two different clusters are only classically correlated, and, in fact for two sites belonging to different clusters ($l\in c$,  $m \in c'$, and $c\neq c' \nonumber$):
\begin{eqnarray}
g_Q(l,m) &=& G_Q(l,m) = 0 ~~~~~~~ \text{(lattice bosons)} \nonumber\\
C_Q(l,m) &=& 0 \nonumber ~~~~~~~~~~~~~~~~~~~~~~~ \text{(Ising spins)}
\end{eqnarray}
given that imaginary-time fluctuations of different clusters are completely uncorrelated. Put differently, the QMF approximation amounts to confuse the total field correlations between two clusters with the thermal part only.

\subsection{Rationale of the QMF approximation}

The separation between thermal and quantum correlations provided in the previous section is not simply a formal operation, given that the thermal and quantum correlation functions (\emph{e.g.}, $g_T$ and $g_Q$) have: 1) widely different magnitudes; and, most importantly 2) widely different spatial structures. 

\subsubsection{Separation in magnitude between quantum and thermal correlations}

From the definition Eq.~\eqref{e.classcorr} one can show \cite{MalpettiR2016} that thermal correlations vanish as $T\to 0$, so that $g_Q = g$ in that limit (and the same identity holds for all correlation functions). On the other hand, it is also obvious that $g_Q \to 0$ (as well as all quantum correlation functions) when $T\to \infty$. Therefore the thermal and quantum correlation functions can be spaced by several orders of magnitude, and in particular $g_T \gg g_Q$ at high temperature. This immediately leads to the simple remark that the cQMF approximation for a given cluster size should be increasingly good as the temperature is raised; while it can be expected to be rather poor in a regime in which quantum correlations nearly saturate the total ones.
At the same time a simple calculation shows that, as $T\to 0$
\begin{eqnarray}
g_T(l,m) &\approx&~ k_B T ~\sum_{n>0} \frac{1-e^{-\beta \Delta E_n}}{\Delta E_n} \langle 0 | b_l^{\dagger} | n \rangle \langle n | b_m | 0\rangle \nonumber \\
&\myeq &  \frac{k_B T}{\Delta E_1} ~ \langle 0 | b_l^{\dagger} | 1 \rangle \langle 1 | b_m | 0\rangle \nonumber
\end{eqnarray}
where $|0\rangle$ is the (grand-)Hamiltonian ground state with $N$ particles, and $|n\rangle$ ($n>0$) are the excited Hamiltonian eigenstates with $N-1$ particles, with energy gaps $\Delta E_n$. 
In the case of Ising spins a similar calculation leads to
\begin{eqnarray}
C_T(l,m) \myeq  \frac{k_B T}{\Delta E_1} ~ \langle 0 | S_l^z | 1 \rangle \langle 1 | S_m^z | 0\rangle \nonumber
\end{eqnarray}
where $|0\rangle$ is again the ground state, while $|1\rangle$ is the first excited state.
 
From the above expressions we see that thermal correlations are suppressed only when the temperature becomes much lower than the lowest excitation gap $\Delta E_1$. For quantum Ising spins with a finite size, the excitation gap is typically always finite except at the critical field $\Gamma_c$ separating a magnetically ordered phase ($\Gamma < \Gamma_c$) from a quantum paramagnetic phase ($\Gamma > \Gamma_c$) \cite{quantumIsingbook}. At the critical point, the gap vanishes with system size as $L^{-z}$ with $z=1$. The situation is even more serious in the case of quantum models with a ground state breaking a continuous symmetry, such as superfluids in dimensions $d=2$ and higher; in that case the lowest energy gap scales as $L^{-d}$, according to the characteristic scaling of the so-called Anderson's tower of states \cite{Anderson1952}. Therefore, in gapless systems the limit $g_Q \to g$ (for lattice bosons) and $C_Q \to C$ is attained at extremely low temperatures, while elsewhere the thermal correlations remain sizable. 

\subsubsection{Length-scale separation between quantum and thermal correlations}

\begin{figure}[ht!]
\includegraphics[width = \columnwidth]{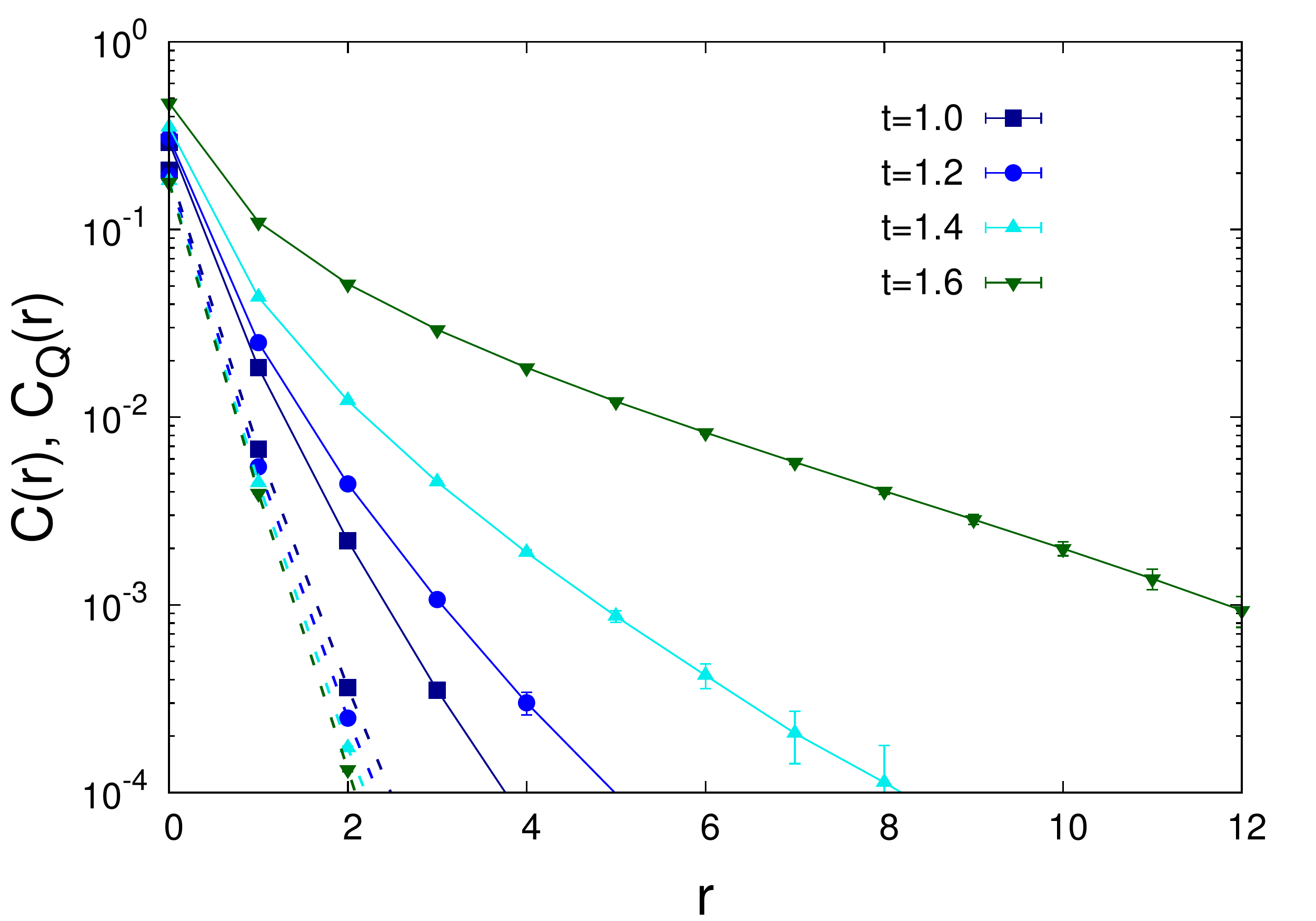}
\caption{Correlation function for the 2$d$ quantum Ising model with $\gamma = 2$ and $L=36$ at various temperatures. Solid lines indicate the total correlations $C(r)$, which are estimated on a finite-size system as 
$C(r) = \langle \sigma^z_i \sigma^z_{i+r}\rangle -  \langle \sigma^z_i \sigma^z_{i+L/2}\rangle$, given that $\langle \sigma_i^z\rangle = 0$.  The dashed lines indicate the quantum correlation function $C_Q(r)$.}
\label{f.corrIsing}
\end{figure}

Quantum correlations at finite temperature are not only suppressed in overall magnitude - as discussed in the previous section - when increasing the temperature, but they display a \emph{different spatial structure} than total (and thermal) correlations. Indeed, as shown in Ref.~\cite{MalpettiR2016} -- and further exemplified in Figs.~\ref{f.corrIsing} and \ref{f.qcorr} -- the generic decay of quantum correlation functions at finite $T$ is \emph{exponential}, with a characteristic quantum coherence length $\xi_Q$ which is completely independent of the correlation length $\xi$ for the decay of the total correlations. This implies that, for generic systems at any finite temperature, thermal and quantum correlations live on \emph{different} length scales, with $\xi > \xi_Q$. This separation in length scales between thermal and quantum correlations represents the most fundamental rationale for the cQMF approximation: indeed the latter introduces a spatial cutoff in the quantum correlations over a length scale $l_c/2$ (the maximum distance between the bulk of the cluster and its boundaries), which would be \emph{a priori} completely arbitrary if quantum correlations did not possess the above-mentioned exponential decay. Therefore the quality of the cQMF scheme is controlled by the ratio $\vartheta = 2\xi_Q/l_c$, and under the condition $\vartheta \ll 1$ the approximation becomes extremely accurate.
 
 \begin{figure}[ht!]
\includegraphics[width = \linewidth]{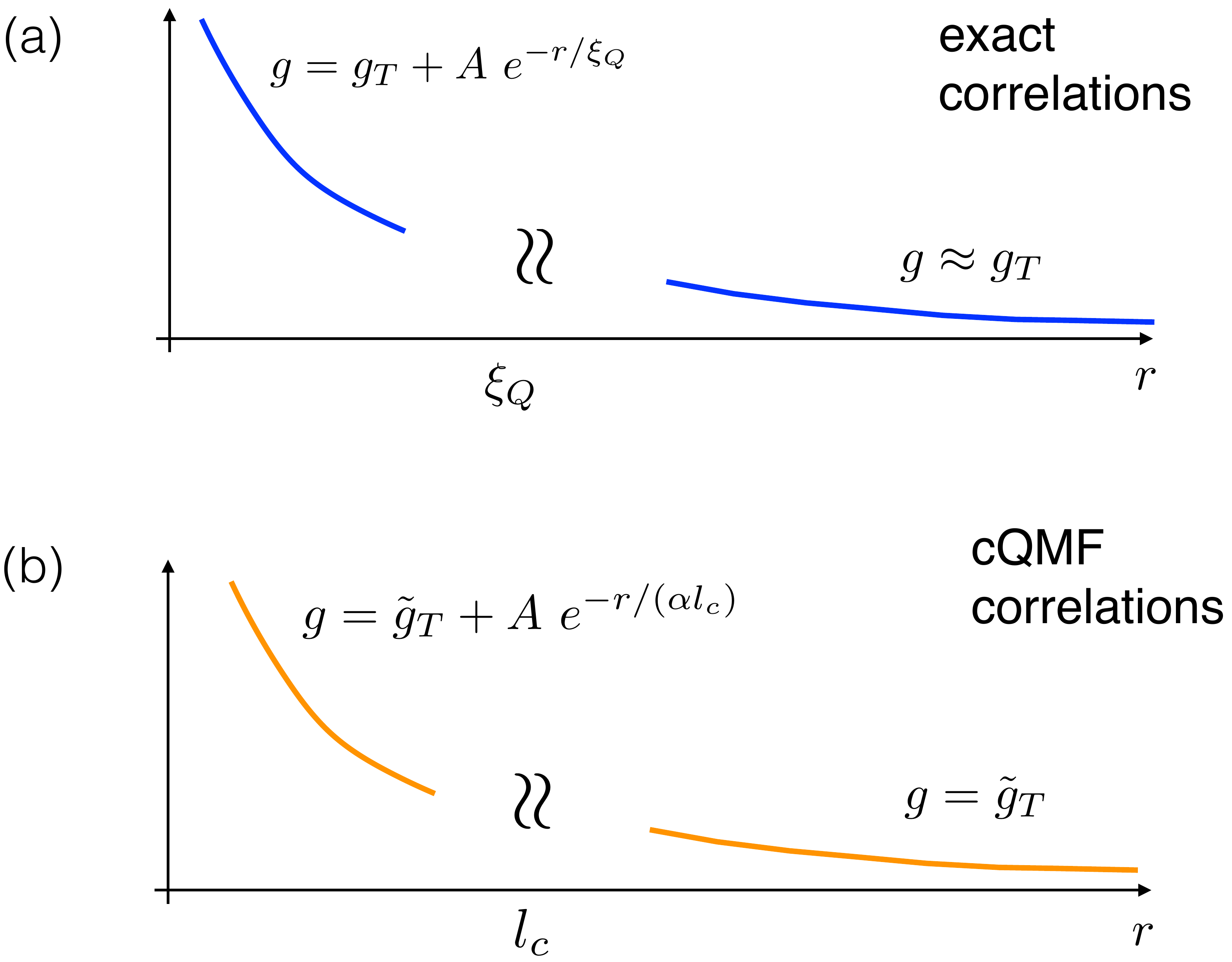}
\caption{(a) Sketch of the correlations in quantum many-body system at finite temperature. Beyond a characteristic quantum coherence length $\xi_Q$ the total correlations ($g$) are very close to the thermal part ($g_T$);
(b) Sketch of the cQMF account for correlations in quantum many-body systems at finite temperature. Beyond a length proportional to the cluster size ($\alpha l_c$) the total correlations are described as \emph{identical} to the (approximate) thermal correlations $\tilde g_T$.}
\label{f.corr}
\end{figure}  
 
  The existence of a finite quantum coherence length $\xi_Q$, beyond which thermal and total correlations are nearly identical (see sketch in Fig.~\ref{f.corr}(a)), implies that for $r \gg \xi_Q$ correlations in a quantum many-body system can be regarded -- to within a very good approximation -- as those generated by an effectively classical system. The long-range ($r \gg \xi_Q$) correlations are supposedly captured by an effective classical model whose local degrees of freedom (on the scale $\xi_Q$) are renormalized in their effective couplings by short-range quantum fluctuations. The cQMF approach is precisely a systematic way of building effective classical theories which describe short-range quantum correlations up to a scale proportional to the cluster linear size $l_c$, and which then identify the total correlations with (approximate) thermal ones ($\tilde g_T$) beyond that scale (see sketch in Fig.~\ref{f.corr}(b)). 
  
 How well can one expect the cQMF approximation to capture the correlations of a quantum system?  
 The general asymptotic form of correlation functions reads  
  \begin{equation}
  A \frac{e^{-r/\xi}}{r^{d-2+\eta}}
  \end{equation}
with temperature- and size-dependent parameters $A$, $\xi$ and $\eta$. The cQMF adds to the above dependencies a further cluster-size dependence, namely $A=A(\vartheta;T,L)$, $\xi=\xi(\vartheta;T,L)$, $\eta=\eta(\vartheta;T,L)$. Being restricted to quantum fluctuations, the cQMF guarantees to capture the correct divergence of the correlation length at a thermal transition, $\xi \sim |T-T_c|^{\nu}$, as well as the correct exponent $\eta$ at the critical point; yet the position ($T_c$) of the critical point  itself may depend significantly on the $\vartheta$ parameter, namely $T_c = T_c(\vartheta)$. In general $T_c(\vartheta>0) > T_c(0)$, namely the cQMF approach overestimates the transition points by systematically underestimating quantum effects. 
  
  As for the $\vartheta$ dependence of the $A$, $\xi$ and $\eta$ parameters of the correlation function, we can make the following, general remarks. If $\xi \sim \xi_Q$, the short-range properties of the correlation function are dominated by quantum effects, and therefore $\xi$ will depend very strongly on $\vartheta$: therefore only the condition $\vartheta \ll 1$ guarantees the faithful reconstruction of the correlation length. On the other hand, when $\xi \gg \xi_Q$ thermal and quantum correlations acquire a genuine separation of scales: one can hope to reconstruct faithfully the long-range aspects of thermal correlations, while only partially accounting for quantum correlations. In particular, in the presence of an extended critical phase with $\xi = \infty$ (such as in the superfluid regime of two-dimensional bosons), the cQMF approximation can ``trivially" capture the correct correlation length $\xi = \infty$ by featuring a critical phase in the same temperature range. Beyond that, it can capture extremely well the $\eta$ exponent (governing the decay of correlations) even for $\vartheta \sim 1$. The main deviation from the exact result is observed in the value of $A$, which is sensitive to the short-range physics in that it contains the quantum renormalization of the effective classical degrees of freedom whose correlations are being probed. Under the above conditions, the tail of the correlation function is well described up to a global prefactor, namely $\tilde g_T \approx B(\vartheta) g_T$, with $B(\vartheta\to 0) \to 1$.

   In the next section we show explicitly the ability and limitations of the cQMF to capture the correlation functions of two paradigmatic quantum models -- the transverse-field Ising model on a square lattice, and the quantum-rotor model on the same lattice -- which lend themselves to a path-integral treatment, and therefore to an application of the cQMF approximation for arbitrary cluster sizes. This allows in turn to test systematically the convergence of relevant physical observables upon increasing the cluster size.

 \section{Two-dimensional transverse-field Ising model and QMF approximation}
 \label{s.Ising_QMF} 
 
 We begin our discussion of the accuracy of the cQMF approximation with the study of the two-dimensional Ising model in a transverse field, namely $J_{ij} = J \delta_{\langle ij \rangle}$, where $\delta_{\langle ij \rangle}$ is a Kronecker delta selecting only nearest-neighboring sites. The physical properties of the system are fundamentally controlled by the ratio $\gamma = \Gamma/J$; $\gamma_c = \Gamma_c/J = 3.04...$ \cite{DengB2002} represents the quantum critical point. 
 
 The action of the system within the cQMF approximation as in Eq.~\eqref{e.cQMFIsing}, as well as the exact action in Eq.~\eqref{e.SI}, lend themselves straightforwardly to a path-integral quantum Monte Carlo (PIMC) study. In particular, the choice of the length $M$ of the Trotter dimension should be made such that the error introduced by the Trotter discretization becomes negligible compared to the statistical Monte Carlo error. Introducing the parameter $\epsilon = \beta \Gamma/M$, controlling the form of the effective classical action in Eq.~\eqref{e.SI} ($K_{ij} = (J/\Gamma) \delta_{\langle ij \rangle} \epsilon $ and $K_{\tau} = (1/2) |\log \tanh \epsilon|$), we generically find that a value $\epsilon = 10^{-2}$ fulfills the above condition (statistical error overcoming the Trotter error) for the spin-spin correlation function $C(l,m)$.
 
 From a technical point of view, the all-to-all couplings introduced by the cQMF approximation (see Fig.~\ref{f.QMF}) prevent one from using the powerful Wolff cluster update scheme for classical spins \cite{Wolff1989} in its standard formulation -- which is best adapted to nearest-neighbor couplings, and which is used whenever $l_c = L$ (namely in the case of the exact numerical treatment). Nonetheless, the time-like couplings retain the nearest-neighbor structure even under the QMF approximation, and a modified, ``time-like" Wolff algorithm can be formulated which, starting from a seed site $(i,k)$, grows one-dimensional clusters in imaginary-time only; and it later accepts or rejects the cluster update on the basis of the energy change in the space-like couplings. Similarly, one can formulate a ``space-like" Wolff algorithm in which all the Ising spins with the same site label $i$ (and forming a chain in imaginary time) are treated as a single site, and updated together: a cluster of chains is then built ``\emph{\`a la} Wolff". We find that the use of both modified Wolff algorithms (the time-like one and the space-like one) is crucial to obtain the correct convergence of the simulations. 

 \begin{figure}[ht!]
\includegraphics[width = 0.85\columnwidth]{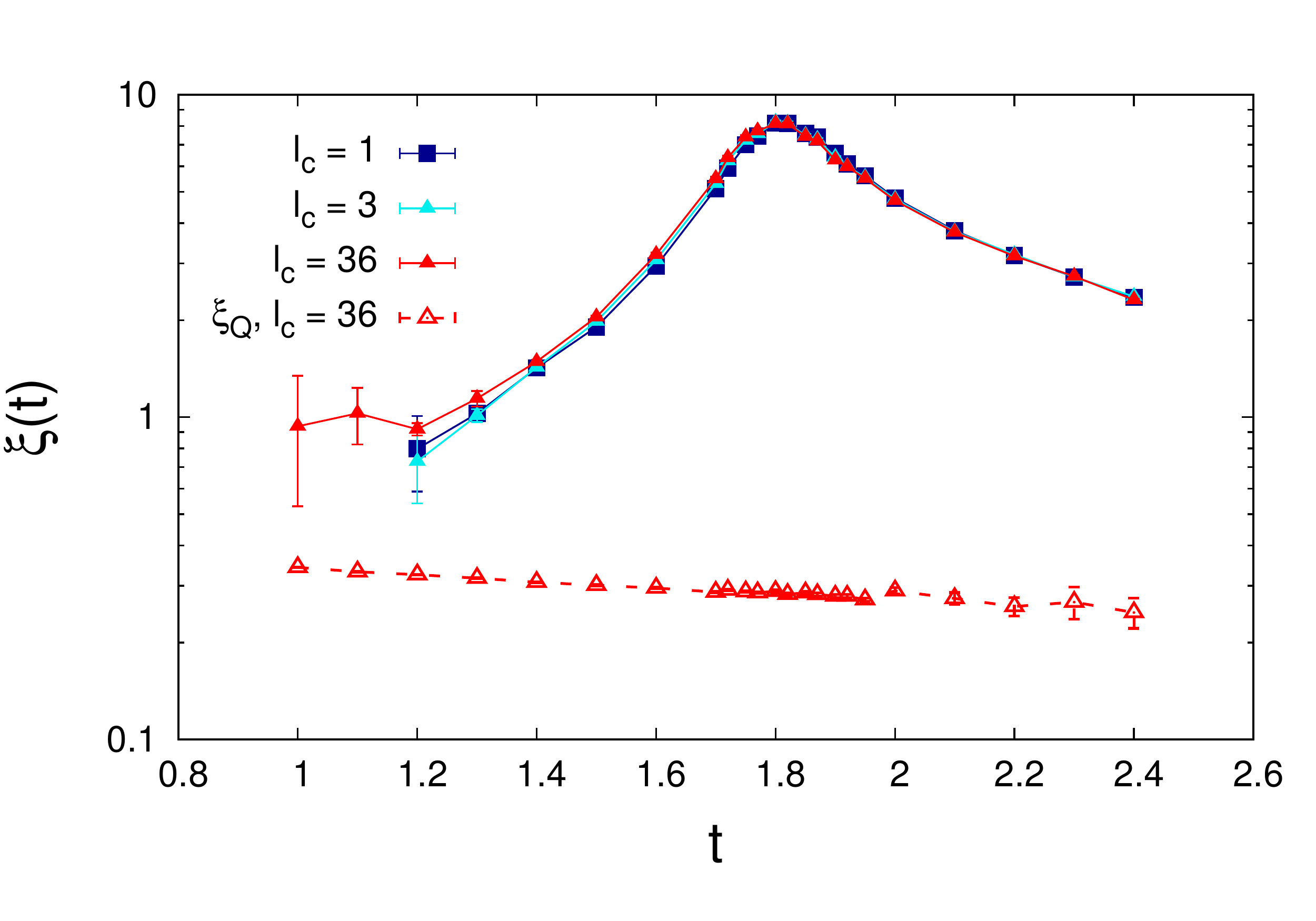}
\includegraphics[width = 0.85\columnwidth]{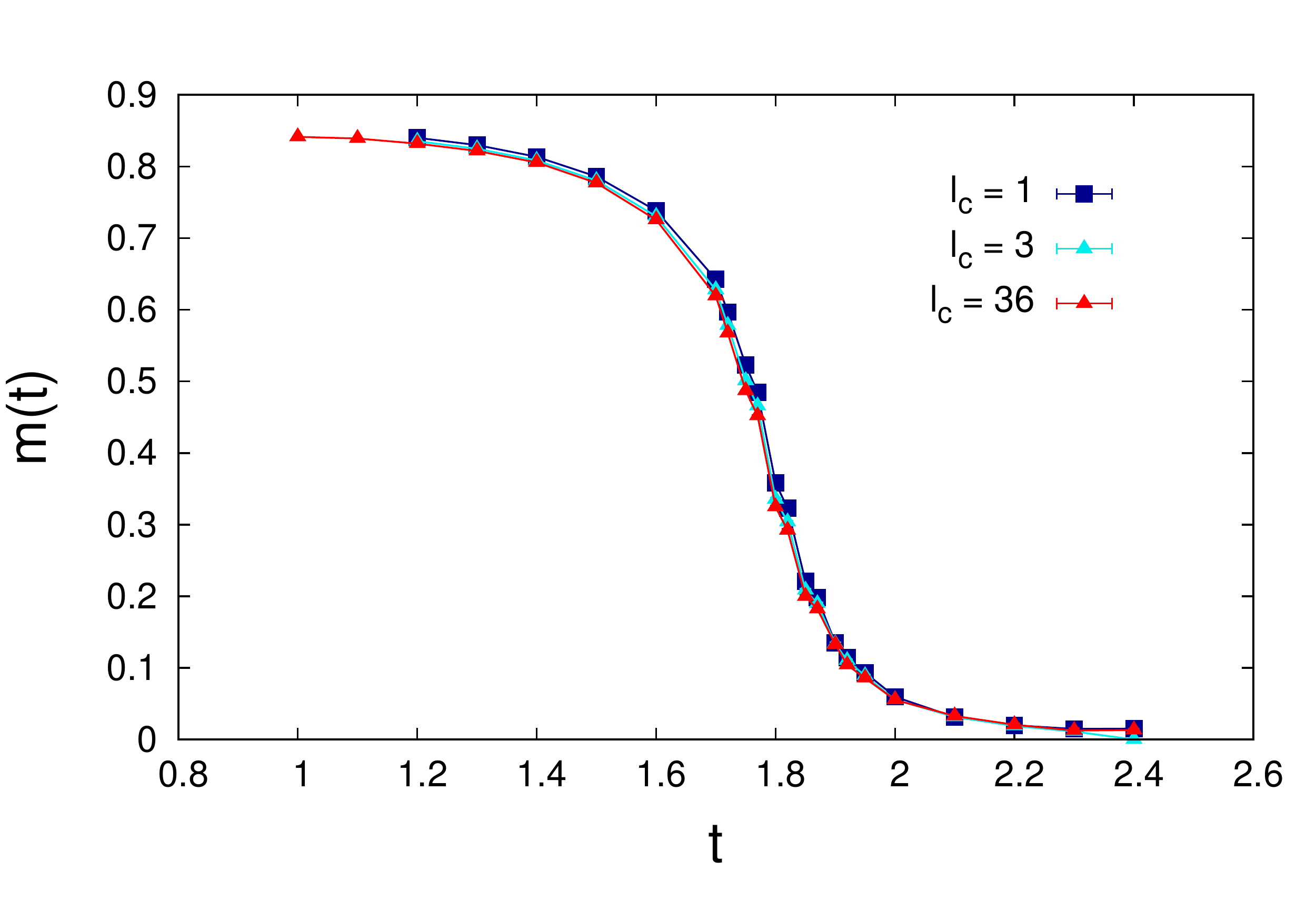}
\caption{Temperature dependence of the correlation length $\xi$ (upper panel) and of the magnetization $m$ (lower panel) for the 2$d$ quantum Ising model with $\gamma = 2$, within the cQMF approximation with variable cluster size $l_c$. The $\xi$ and $m$ parameters are extracted from fits of the correlation function $\frac{1}{L^2} \sum_i \langle \sigma_i^z \sigma_{i+r}^z \rangle$ to Eq.~\eqref{e.fit}. We also show the quantum coherence length $\xi_Q$, extracted from similar fits to the numerically exact quantum correlation function $C_Q$. All the data are for $L=36$.}
\label{f.mxigam2}
\end{figure}   
 
 \begin{figure}[ht!]
\includegraphics[width = 0.85\columnwidth]{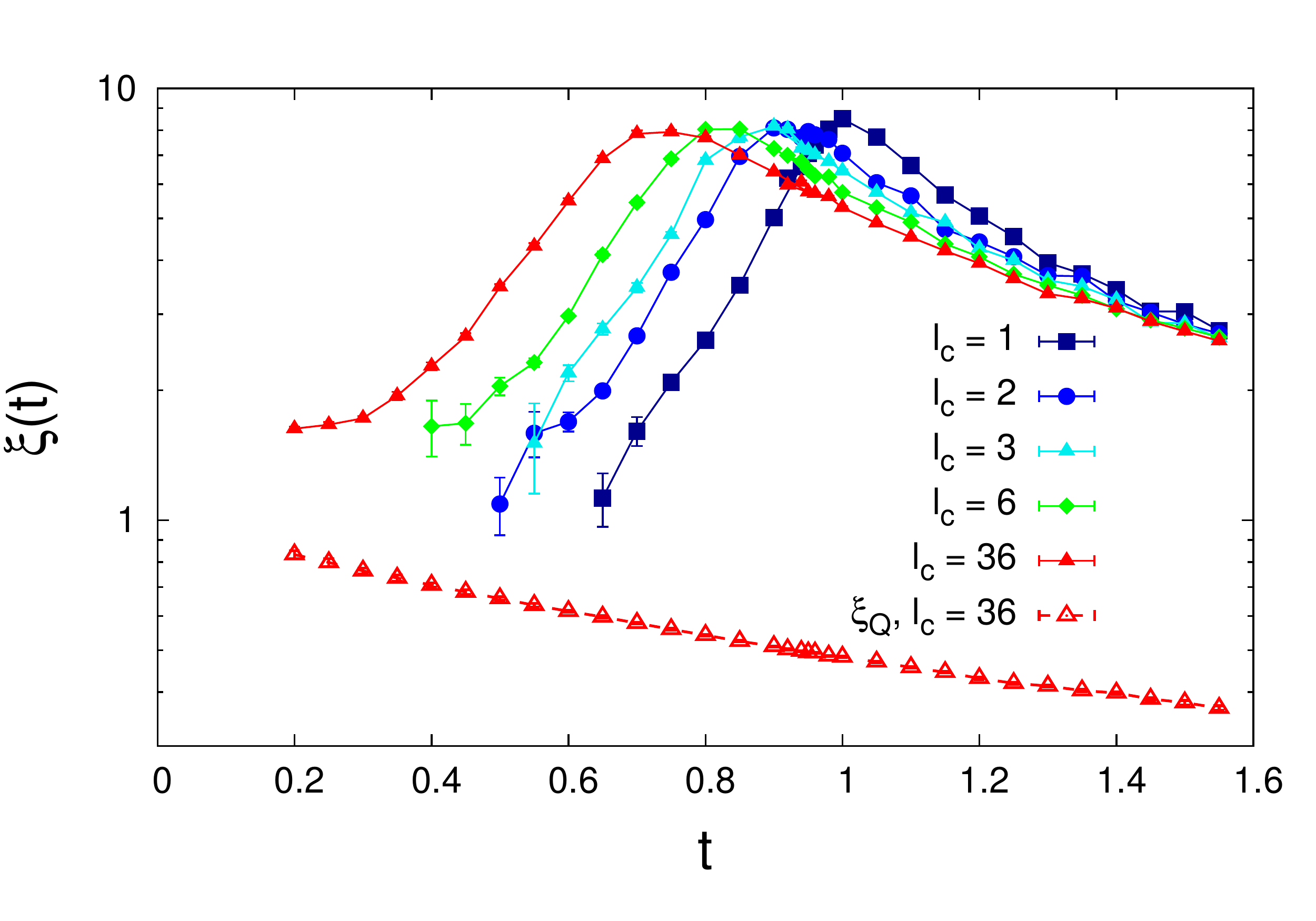}
\includegraphics[width = 0.85\columnwidth]{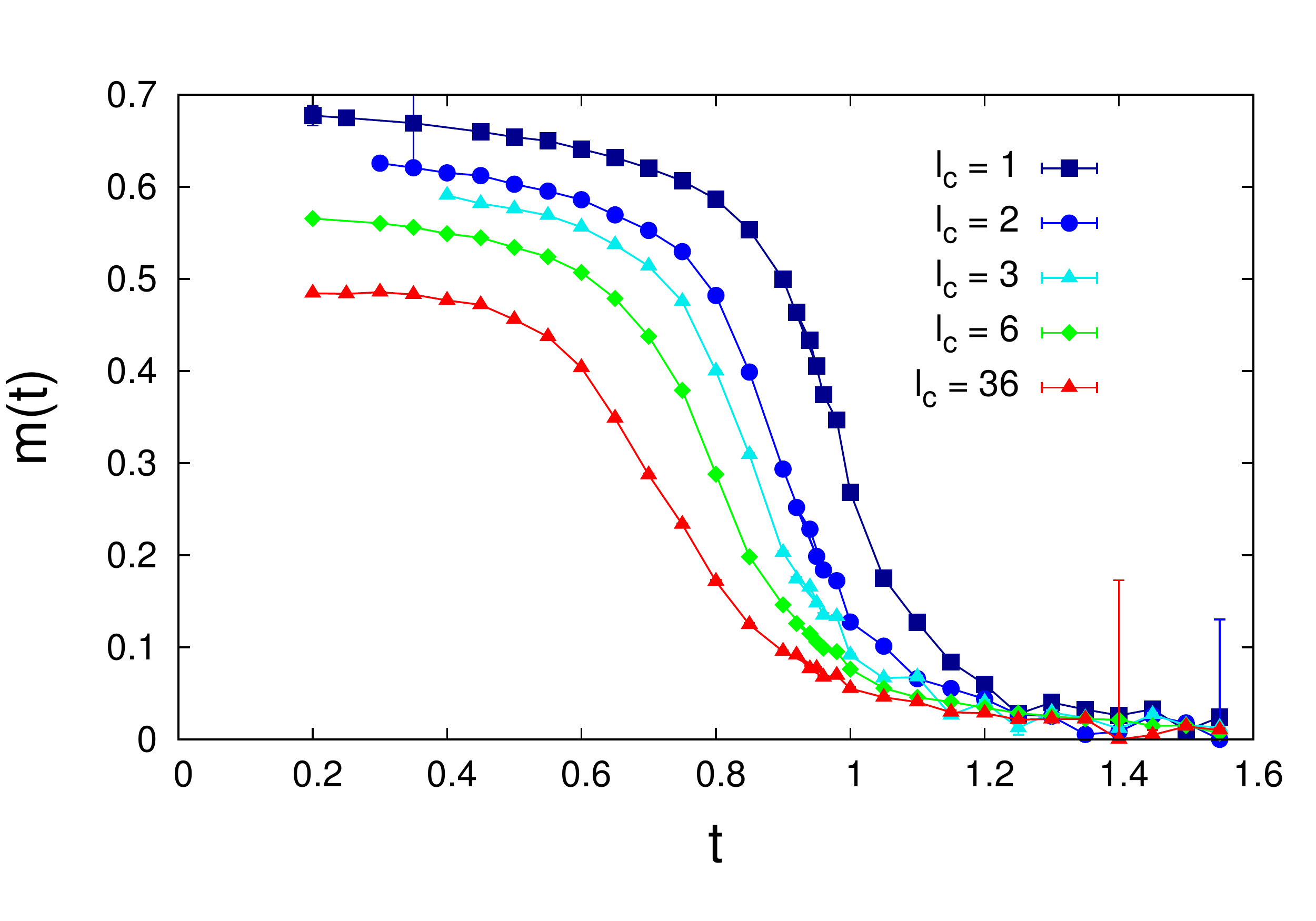}
\caption{Temperature dependence of the correlation length $\xi$ (upper panel) and of the magnetization $m$ (lower panel) for the 2$d$ quantum Ising model with $\gamma = 2.9$, within the cQMF approximation with variable cluster size. Fit procedure and other data parameters as in Fig.~\ref{f.mxigam2}.}
\label{f.mxigam2.9}
\end{figure}    
 
 \subsection{Description of correlation functions: different regimes}
 
In the following we focus our attention on the total ($C(r)$) and quantum ($C_Q(r)$) correlation function for quantum Ising spins for cQMF calculations with increasing cluster size. We consider $L\times L$ square lattices with periodic boundary conditions, paving the lattice with square $l_c\times l_c$ clusters such that $L$ is an integer multiple of $l_c$. 
Fig.~\ref{f.corrIsing} shows these two correlation functions for the 2$d$ transverse-field Ising model with $\gamma = 2$ at different temperatures approaching from below the critical temperature $t_c \approx 1.720(5)$. One clearly observes that, while the total correlation function increases its range when approaching $t_c$ (becoming algebraically decaying as $t=t_c$) the quantum one shows an increasingly marked exponential decay with a very small quantum coherence length ($\xi_Q < 1$). Hence the long-range tail of the correlation function should be captured quantitatively at the semiclassical level in this regime. 
 
 We perform a systematic comparison between the numerically exact results for the correlation function of quantum Ising spins and those stemming from the cQMF approximation by fitting the non-connected correlation function $\tilde C(r) = \langle \sigma^z_i \sigma^z_{i+r} \rangle$ on a finite size $L$ with the $L$-periodic form 
 \begin{equation}
A~\left [ \frac{e^{-r/\xi}}{r^{\eta}} + \frac{e^{-(L-r)/\xi}}{(L-r)^{\eta}} \right ] + m^2~.
\label{e.fit}
\end{equation}
Here, as it is customarily done, we extract a finite-size estimate of the order parameter $m$ via the asymptotic value of the non-connected correlation function. 
 Moreover we have set the $\eta$ exponent of algebraic decay to its expected value at $t_c$, $\eta = 1/4$, in order to reduce the fitting parameters to $A$, $m$ and $\xi$. The quantum correlation function is also fitted to a similar form but setting $m$ and $\eta$ to zero. The fits have been performed for $L=36$ and for distances $r \geq 5$ (for the total correlation function) and $r\geq 1$ (for the quantum correlation function). Figs.~\ref{f.mxigam2} and \ref{f.mxigam2.9} show the results of the fits for $\gamma = 2$ and $2.9$ respectively -- the first value of the transverse field induces moderate quantum fluctuations, while the second value is only $7\%$ below the quantum critical point. 
 
 In the case $\gamma = 2$ we observe that the correlation length and order parameter around the critical point are very well captured already at the lowest level in the cQMF approximation, namely for $l_c=1$. Therefore one can conclude that the thermal critical behavior of the 2$d$ quantum Ising model for $\gamma= 2$ is well reproduced by an effective classical model in which all quantum effects are contained as a renormalization of the local spin, namely via the effective action
 \begin{equation}
 S_{\rm eff}^{(l_c=1)} = - K \sum_{ij} \bar \sigma_i \bar \sigma_j - K_{\tau} \sum_i \sum_k \sigma_{i,k} \sigma_{i,k+1} 
 \label{e.SeffIsing}
 \end{equation}
 corresponding to the cQMF approximation for $l_c=1$.
 This action takes the form of a classical Ising model for time-averaged spins $\bar\sigma_i$; the average is taken over an imaginary-time evolution which is governed by the $K_{\tau}$ term, providing the stiffness to imaginary-time fluctuations. 
 The success of the cQMF approach in its lowest order can be understood by inspecting the value of the quantum coherence length $\xi_Q$: throughout the temperature range shown in Fig.~\ref{f.mxigam2} one has $\vartheta \lesssim 0.6$ for $l_c=1$. 
 
 In the case $\gamma = 2.9$, on the other hand, we observe a strong cluster-size dependence of the estimates of the correlation length and order parameter in the vicinity of the thermal transition. In the case of the smallest cluster ($l_c=1$) the condition $\vartheta \ll 1$ is only reached well above the transition ($\vartheta > 0.6$ throughout the temperature range of Fig.~\ref{f.mxigam2.9}). For larger clusters, the condition $\vartheta \ll 1$ can instead be fullfilled, and indeed the results for $\xi$ converge towards the exact ones upon increasing $l_c$ when that condition is satisfied -- \emph{e.g.} for $l_c=6$ in the highest temperature range of Fig.~ \ref{f.mxigam2.9}.  
 
 The comparison between the results at $\gamma = 2$ and $\gamma = 2.9$ indicates that a sharp value of the $\vartheta$ parameter for convergence of the cQMF result cannot really be fixed -- as we shall see in the following, the convergence towards the exact result is actually algebraic in the cluster size. In this spirit, a consideration is in order with regards to the cluster-size scaling, which shall be useful also to the subsequent discussion. Among all cluster decompositions of the lattice used in the cQMF approximation, only the one with $l_c=1$ does not break translational invariance. The other decompositions, on the other hand, introduce an artificial distinction between bulk and boundary sites in the cluster, and treat the correlations involving sites of different types in a very different manner. Indeed, for bulk sites sitting at a distance $r$ from the boundaries, with $\xi_Q < r \leq l_c/2$, the quantum correlations can be potentially described faithfully; on the other hand, for all other sites quantum correlations are described much more poorly. 
 
 For systems with periodic boundary conditions (such as the ones studied here), translational invariance can be restored by symmetrizing the density matrix over all possible cluster translations, that is to say, by modifying the cQMF Ansatz Eq.~\eqref{e.rhoQMF} to the form 
 \begin{equation}
 \hat\rho_{\rm cQMF} = \frac{1}{N_{\cal T}} \sum_{\cal T} \sum_{\{\Psi\}} p(\{\Psi\}) \left [\otimes_c ~\hat\rho_{_{{\cal T}(c)}} (\{\Psi\}) \right] 
 \end{equation}
 where ${\cal T}$ are the independent translations of the clusters (which are in a number $N_{\cal T} = l_c^2$ for $l_c\times l_c$ square clusters), and ${\cal T}(c)$ is the corresponding translated cluster. Here $\sum_{\{\Psi \}}$ is the short-hand notation for the integral over the boundary auxiliary fields appearing in Eqs.~\eqref{e.rhoQMF} and \eqref{e.rhoQMF_2}. When calculating the correlation function,  the above symmetrization simply amounts to averaging correlations between points $i$ and $i+r$ on all the translations of the reference point $i$. Restoration of translational invariance, while necessary to obtain physically meaningful results, degrades the quality of the description of quantum correlations by mixing together results with bulk as well as boundary reference points.

  \begin{figure}[ht!]
\includegraphics[width = \linewidth]{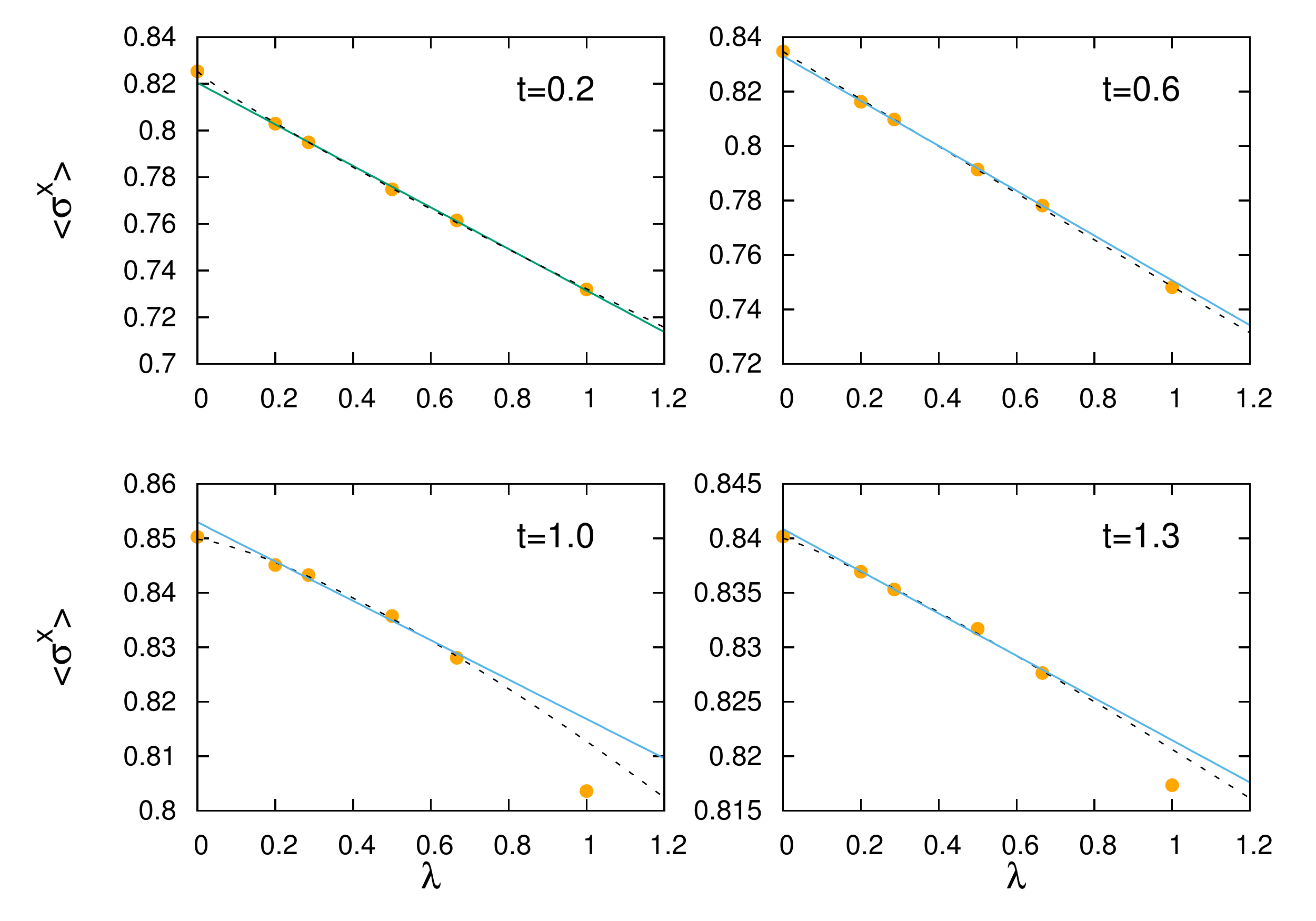}
\caption{Scaling of the transverse magnetization $\langle \sigma^x \rangle$ with the cluster parameter $\lambda$ for quantum Ising spins on the square lattice with $\gamma = 2.9$ and $L=36$.  $\lambda$ values correspond to cluster linear sizes $l_c =$ 1, 2, 3, 6, 9 and 36 (in decreasing order). The solid and dashed lines are a linear fit (excluding the points at $\lambda = 0$ and 1) and a power-law fit $a_1+a_2*\lambda^{a_3}$ (excluding the point at $\lambda = 1$), respectively. }
\label{f.sxscaling}
\end{figure}

  \begin{figure}[ht!]
\includegraphics[width = \linewidth]{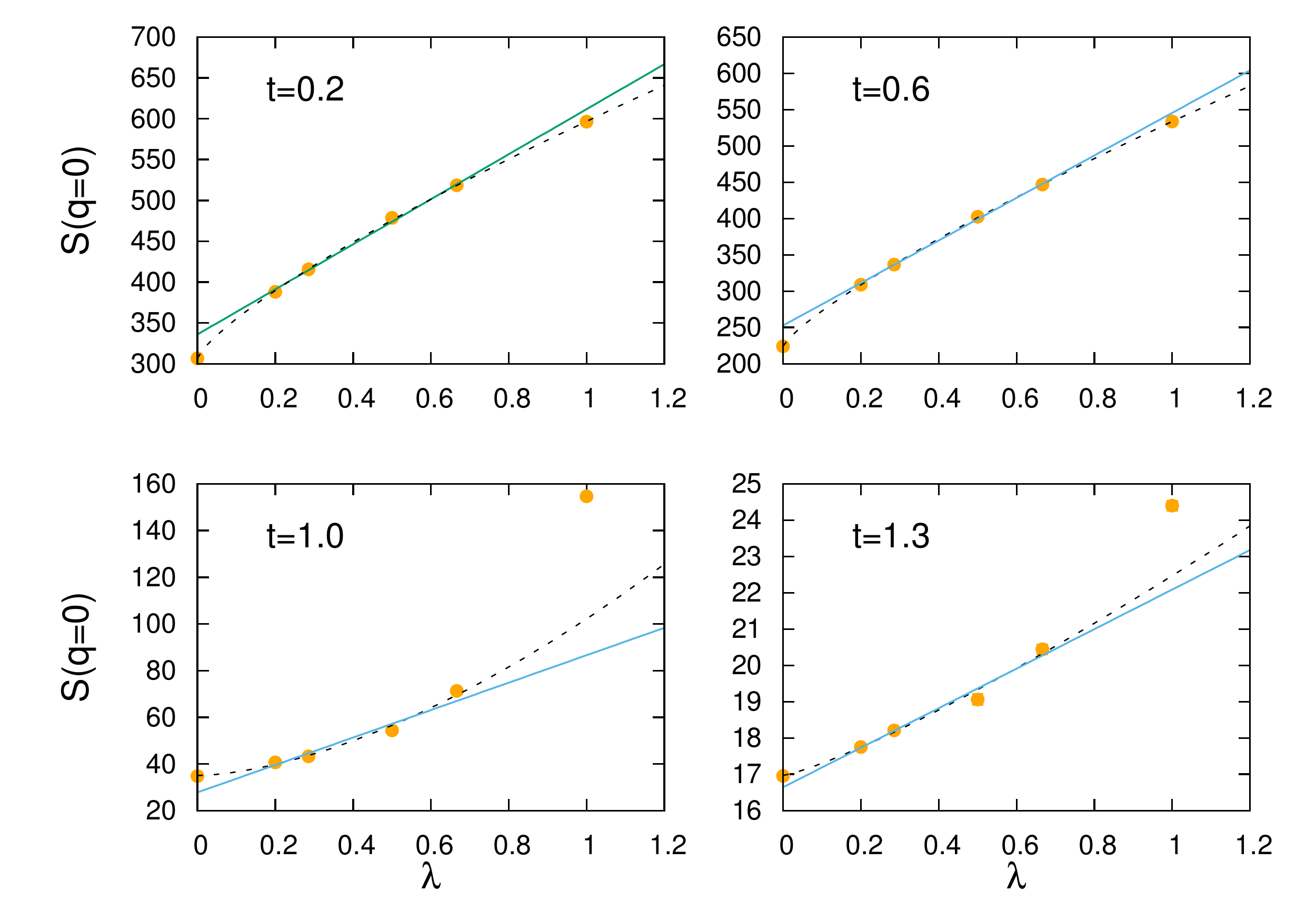}
\caption{Scaling of the structure factor peak $S({\bm q}=0)$ with the cluster parameter $\lambda$ for quantum Ising spins on the square lattice with $\gamma = 2.9$ and $L=36$. Lines and parameters as in Fig.~\ref{f.sxscaling}.}
\label{f.Sq0scaling}
\end{figure}  
       
 \subsection{Cluster-size scaling} 
 
  Given the ability of path-integral Monte Carlo to reconstruct the thermodynamics of the quantum Ising model under the cQMF approximation with arbitrary size of the clusters, we can perform a systematic analysis of the convergence of the results of the cQMF approximation upon increasing the cluster linear size $l_c$. Figs.~\ref{f.sxscaling} and \ref{f.Sq0scaling} show the cluster-size scaling of the transverse magnetization $\langle \sigma^x \rangle$ and of the peak in the structure factor, $S(\bm q=0) = L^{-2} \sum_{ij} \langle \sigma_i^z \sigma_j^z \rangle$, probing the local properties and the non-local correlations of the system respectively.  Not surprinsingly, we find that the cQMF approximation systematically underestimates the transverse magnetization and overestimates the structure-factor peak (namely the longitudinal correlations), meaning that it systematically underestimates the effect of the transverse field in magnetizing the system and in disrupting the spontaneous correlations. When plotted as a function of the boundary-to-bulk ratio $\lambda$, the data nicely fall onto what appears as a simple power-law curve, namely, given the generic observable $O$, one has a rather good fit with a form $O(\lambda) = O_{\rm exact} + a_O \lambda^{b_O}$, where $O_{\rm exact}$ is the exact result in the limit $\lambda \to 0$ and $a_O$ and $b_O$ are fitting parameters depending on the observable in question, as well as on the temperature and the transverse field. In general the exponent $b_O$ is found to grow with temperature, namely the convergence with cluster size is significantly faster at higher temperature.  The figures also show that a linear fit -- which is generally used in cluster MF studies at zero temperature -- may well approximate the results locally, but its extrapolation to $\lambda = 0$ not always recovers the exact value. A power-law fit with variable power is far superior, although it needs three fitting variables which require having access to at least four different cluster sizes. We also observe that the smallest cluster $l_c=1$ usually falls out of the asymptotic power law recovered for $\lambda \to 0$, so it should be eliminated from the data. 
  
  The power-law convergence of the cQMF results with cluster size -- already anticipated in the previous section -- shows that the system lacks a characteristic cluster length scale for convergence. This is in apparent contradiction with the existence of a quantum coherence length $\xi_Q$, which could be naively expected to set the characteristic length beyond which cluster-size convergence is achieved. Yet this expectation misses the crucial observation that clusters are composed of bulk and boundary sites, and that quantum correlations of the latter with the rest of the system are always described poorly because they vanish with at least one nearest neighbor (or two when dealing with corner sites). Hence, even though quantum correlations disappear over distances which are a few multiples of $\xi_Q$, the convergence of cluster-size scaling is actually controlled not by the $\vartheta$ ratio, but rather by the boundary-to-bulk ratio $\lambda$.

 \section{Quantum rotors and QMF approximation}
 \label{s.QR_QMF}
 
\begin{figure}[ht!]
\includegraphics[width = 0.8\columnwidth]{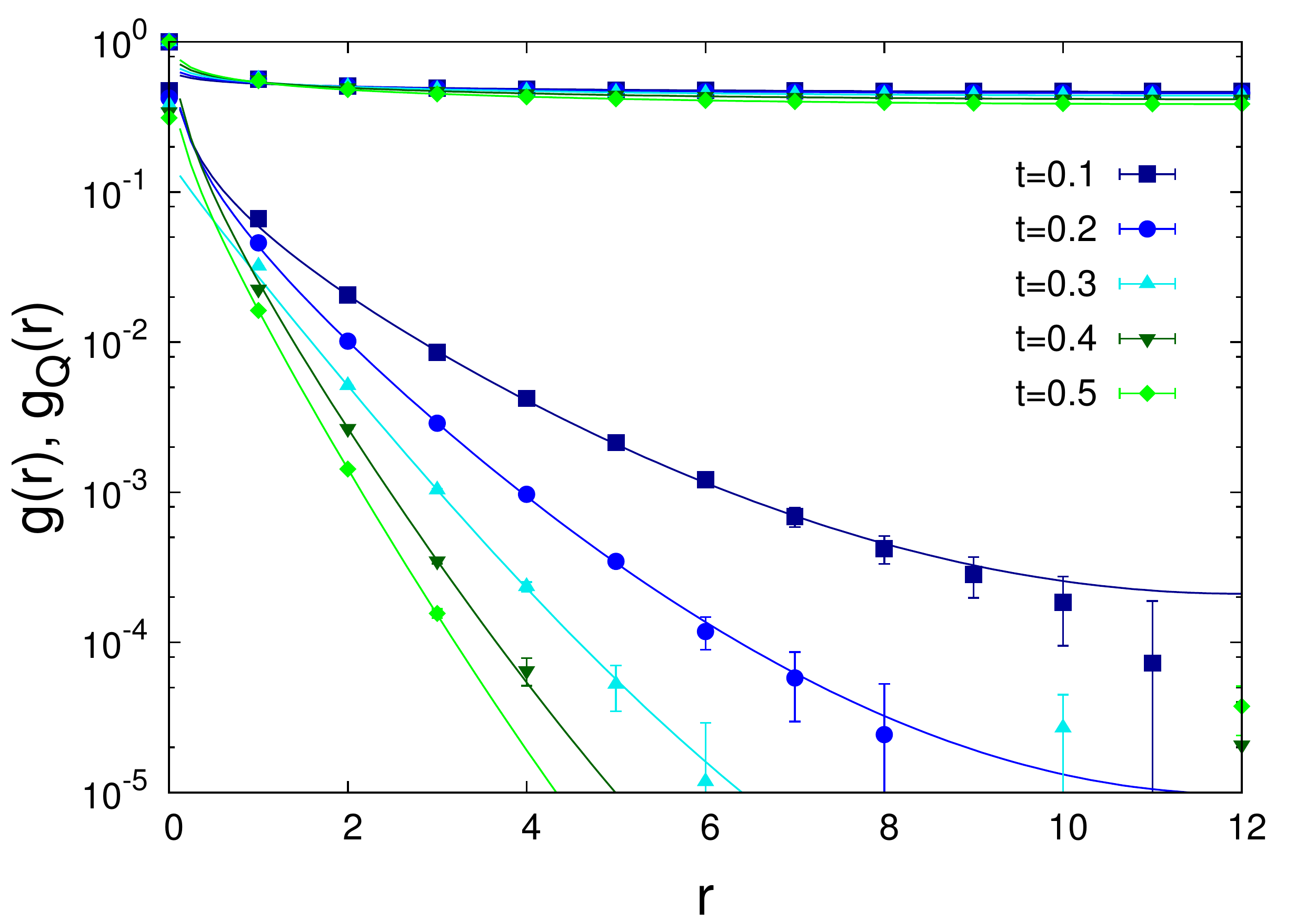}
\caption{Total ($g$) vs quantum ($g_Q$) field correlations for 2$d$ quantum rotors on the square lattice.
Here we consider quantum rotors with $u =U/(2Jn) = 3$ on a $L=24$ lattice for different reduced temperatures  $t = k_B T/(2Jn)$ below the  KT critical temperature $t_{\rm KT} \approx 0.64$. The upper curves refer to the total correlations, while the lower ones to the quantum correlations. Solid lines are algebraic fits $A*d(x|L)^{-\eta}$ for the total correlations, and exponential fits $A'*e^{-d(x|L)/\xi_Q} *d(x|L)^{-\eta'}$ for the quantum correlations, where $d(x|L) = (L/\pi) \sin(\pi x/L)$ is the chord length.}
\label{f.qcorr}
\end{figure}

 In this section we focus our attention on a fundamental lattice-boson theory, namely the quantum rotor model. The latter model is best understood starting from the Bose-Hubbard model \cite{Fisheretal1989}, namely Eq.~\eqref{e.H} with $V_{ij}=0$ and $J_{ij} = J$ for nearest-neighbor bonds $\langle ij \rangle$ and zero otherwise. Up to an additive constant, its Hamiltonian can be rewritten as
 \begin{equation}
 {\cal H} = - J \sum_{\langle ij \rangle} (\hat b_i^{\dagger} \hat b_j + {\rm h.c.}) + \frac{U}{2} \sum_i (\hat n_i - \nu)^2
 \label{e.BH}
 \end{equation}
  where $\nu = \mu/U + 1/2$ is the average density. Taking $\nu$ to be an integer $\delta \hat n_i = \hat n_i-\nu$ can be considered as being an angular momentum operator canonically conjugated to a phase operator $\hat \theta_i$ with commutation relations $[\hat \theta_i, \delta \hat n_i] = i$, so that $\delta \hat n_i = -i \frac{\partial}{\partial \theta_i}$. Moreover, in the limit of very large average filling, $\nu \gg 1$, one can adopt a phase-number decomposition of the Bose operator, $\hat b_i \approx e^{i\hat \theta_i} \sqrt{\hat n_i}$, and neglect number fluctuations in the hopping term of Eq.~\eqref{e.BH}, $\hat b_i \approx \sqrt{\nu}~ e^{i\hat\theta_i}$. This then leads to the quantum-rotor Hamiltonian
 \begin{equation}
 {\cal H}_{\rm QR} = -2J\nu \sum_{\langle ij \rangle} \cos(\theta_i - \theta_j) - \frac{U}{2} \sum_i   \frac{\partial^2}{\partial \theta^2_i}
 \end{equation} 
 where we have dropped the operator notation for $\theta_i$, as we are now working in an explicit phase representation of the Hamiltonian. In summary the quantum-rotor model represents the limit of the Bose-Hubbard model for large, integer filling.  Given the natural energy scale $2J\nu$ appearing in the hopping term, we shall hereafter normalize all other energy scales to this one, namely we introduce the reduced interaction $u = U/(2 J\nu)$ and the reduced temperature $t = k_B T/(2J\nu)$.   
 In the following we consider the explicit case of quantum rotors on the square lattice, which possess a superfluid phase with algebraically decaying correlation function $g$ for temperatures below a critical Kosterlitz-Thouless transition $t < t_{\rm KT}$, as shown in Fig.~\eqref{f.qcorr}. The critical $t_{\rm KT}$  temperature is a decreasing function of the interaction $u$, and it vanishes at the quantum critical point $u_c \approx 5.8$ beyond which the ground state of the system becomes a gapped Mott insulator with exponentially decaying correlations.



 \subsection{Path-integral representation of quantum rotors and QMF approximation}
 \label{s.PIQR}
 
 The path-integral representation of the partition function for the quantum rotor can be naturally achieved using coherent states with unit norm, namely $\phi_i = e^{i\theta_i}$, { representing the eigenstates of the operator $e^{i\hat{\theta}_i}$}.  
 The path-integral expression for the partition function takes then the form 
 \begin{equation}
 {\cal Z}_{\rm QR} = \int {\cal D}[\{e^{i\theta_i(\tau)}\}] ~e^{-S_{\rm QR}[\{e^{i\theta_i(\tau)}\}]}~.
 \end{equation} 
The quantum-rotor action $S_{\rm QR}$ is conveniently expressed upon Trotter-Lie discretization of imaginary time into $M$ slices of length $\delta \tau = \beta /M$, and taking the limit $M\to \infty$.  The imaginary time slice $\delta \tau$ is combined with the repulsion energy to give the ratio $\epsilon = \delta\tau ~U$, which is the (small) parameter fundamentally controlling the quality of the Trotter approximation. In particular for $\epsilon \to 0$ the Villain approximation can be used to reduce the Euclidean action $S_{\rm QR}$ to an  effective classical XY model in ($d$+1) dimensions \cite{Wallinetal1994,Sondhietal1997}
  \begin{eqnarray}
 S_{\rm QR} \approx  \sum_{k=1}^{M} \Big [ &-&K \sum_{\langle ij \rangle} \cos(\theta_{i,k}-\theta_{j,k}) \nonumber \\
 &-& K_{\tau}  \sum_i   \cos(\theta_{i,k}-\theta_{i,k+1}) ~\Big ]
 \label{e.XY}
 \end{eqnarray}
where we have introduced the coupling constants $K = \epsilon/u$ and $K_{\tau} = 2/\epsilon$ for the ``space-like" and ``(imaginary)-time-like" couplings respectively, and the Trotter-discretized phase field $\theta_i(\tau) \to \theta_{i,k}$. 
   
  When applied to the quantum-rotor model, the cQMF approximation amounts to a simple redefinition of the couplings in the effective XY model of Eq.~\eqref{e.XY}. Indeed the cQMF approximation addresses the space-like ($K$) couplings between sites belonging to different clusters, and it redefines them from local in imaginary time to completely non-local -- which is the same as assuming that clusters interact via imaginary-time averaged bosonic fields, $\overline{e^{i\theta_i}}$. The cQMF effective action then breaks up into an intra-cluster part and an inter-cluster part:
  \begin{equation}
 S_{\rm QR} \QMF \sum_c S_c + \sum_{c \neq c'}  S_{cc'}
  \end{equation}
 where $c, c'$ are pairs of interacting clusters. The intra-cluster Hamiltonian is simply the effective Hamiltonian Eq.~\eqref{e.XY} specified to the sites and links within a given cluster $c$:
  \begin{eqnarray}
 S_c =  &\sum_{k=1}^{M} &  \Big [ -K \sum_{\langle ij \rangle, i,j \in c} \cos(\theta_{i,k}-\theta_{j,k}) \nonumber \\
 &&~~- K_{\tau}  \sum_{i \in c}  \cos(\theta_{i,k}-\theta_{i,k+1}) ~\Big ]
 \label{e.XYintra}
 \end{eqnarray} 
  while the inter-cluster Hamiltonian contains the mean-field couplings in imaginary time between neighboring clusters:
  \begin{equation}
 S_{cc'} =  -\frac{K}{M} ~\sum_{k,k'=1}^{M}  ~\sum_{\langle ij \rangle, i \in c, j\in c'} \cos(\theta_{i,k}-\theta_{j,k'})~.
 \label{e.XYinter}
 \end{equation}

A path-integral evaluation of the partition function within the cQMF approximation amounts therefore to simulating a classical XY model with spatially anisotropic couplings and all-to-all couplings in imaginary time for links in between clusters (see again Fig.~\ref{f.QMF} for a cartoon). Analogously to what done in the case of quantum Ising spins, we simulate such a model numerically making use of the PIMC method. Similar considerations on the modifications of the update algorithms, as those made in Sec.~\ref{s.Ising_QMF}, also apply in this case. 


 \begin{figure*}[ht!]
\includegraphics[width = 0.9\linewidth]{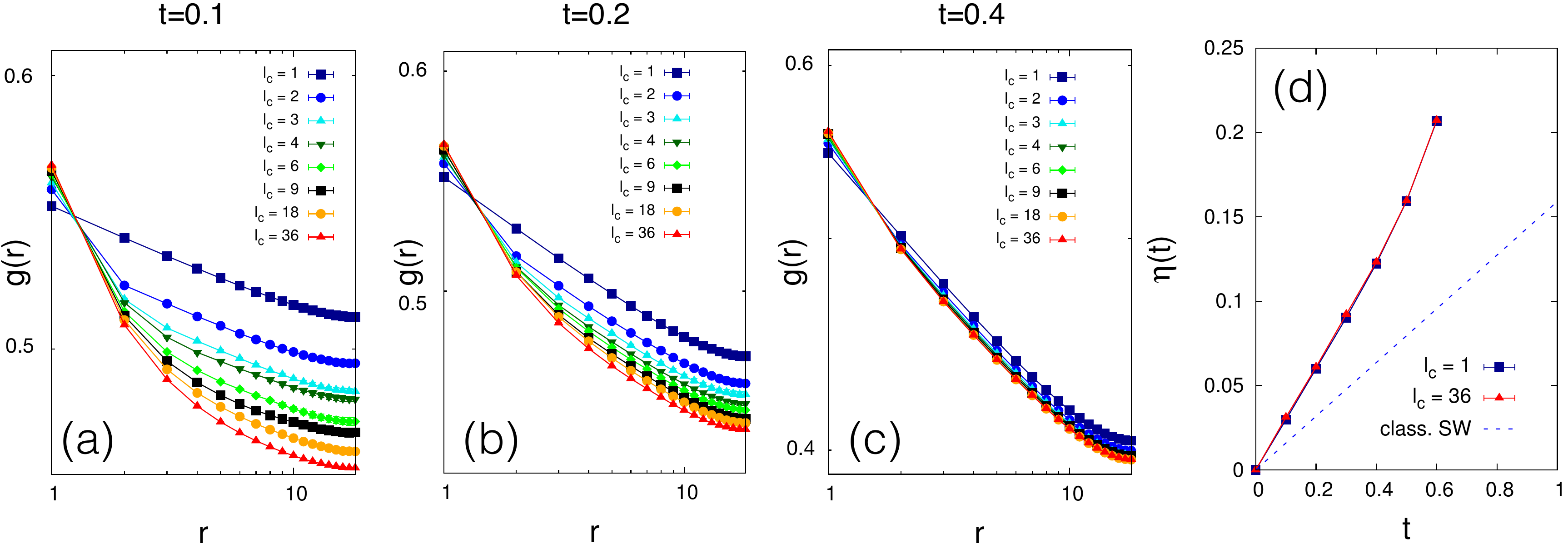}
\caption{(a-c) Total correlations within the cQMF approximation with variable cluster size $l_c$ for quantum rotors on the square lattice. Here $u=3$ and $L=36$. (d) Temperature dependence of the $\eta$ exponent extracted from power-law fits to the correlation function with $l_c=1$ and $l_c = L = 36$ (see text), over the range $r\in [3,L/2]$; the result is contrasted to that of classical spin-wave (SW) theory, $\eta(t) = t/(2\pi)$.}
\label{f.corrQMF}
\end{figure*}

\begin{figure}[ht!]
\includegraphics[width = 0.9\linewidth]{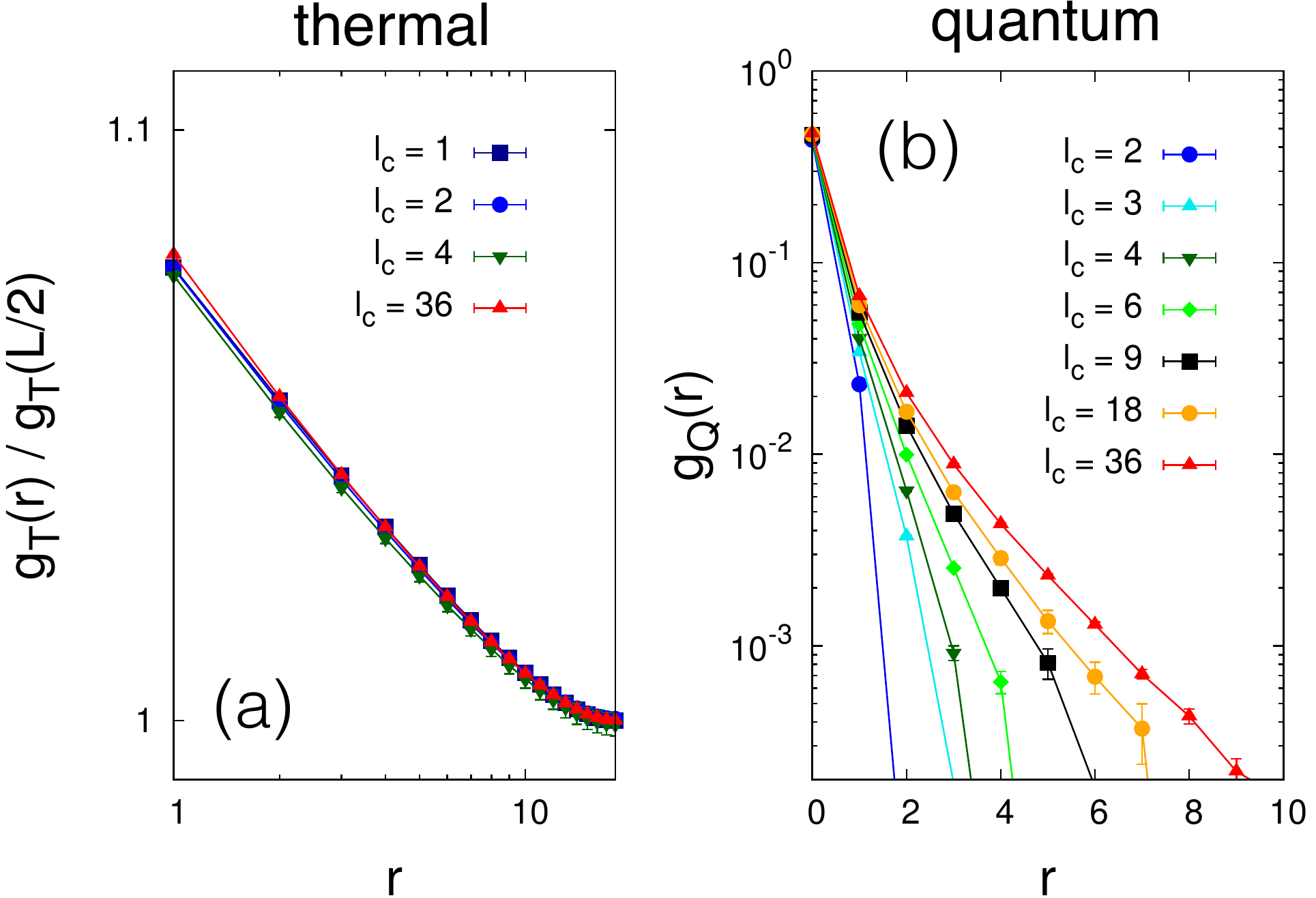}
\caption{Thermal vs. quantum correlations in the superfluid phase of 2$d$ quantum rotors ($t=0.1$, $u=3$, $L=36$) within the cQMF approximation. (a) Thermal correlations $g_T$ for different cluster sizes, normalized to the value at maximum distance $r=L/2$; (b) Quantum correlations $g_Q$ for different cluster sizes.}
\label{f.corr_anatomy1}
\end{figure}

  \begin{figure}[ht!]
\includegraphics[width = 0.6\linewidth]{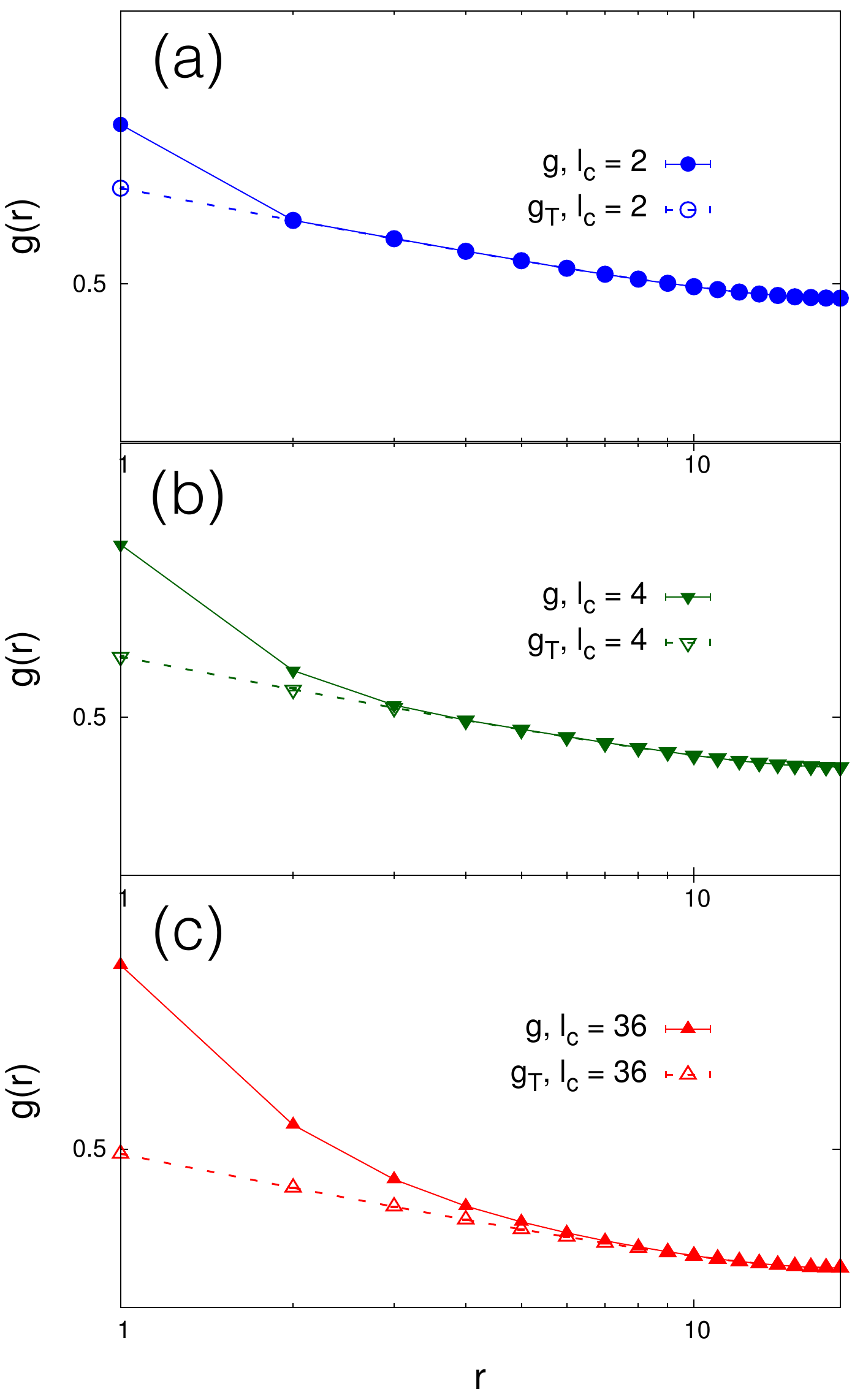}
\caption{Total vs. thermal correlations of  2$d$ quantum rotors ($t=0.1$, $u=3$, $L=36$) for three different cluster sizes. The deviation of total correlations $g$ from a power-law decay at short distance comes entirely from the quantum correlations.}
\label{f.corr_anatomy2}
\end{figure}

\subsection{cQMF results for two-dimensional quantum rotors}

\subsubsection{Description of correlations}

 In the following we discuss our Monte Carlo results showing the structure of correlations in the 2$d$ quantum rotor model, and the ability of the cQMF approach to capture that structure.  
 We shall be primarily concerned with the low-temperature, superfluid phase of quantum rotors, as that phase exhibits critical correlations over a finite temperature range, posing the biggest challenge to any approximate description. Fig.~\ref{f.qcorr} shows that, in the face of critical total correlations ($g$) at finite temperature, the quantum correlation function $g_Q$ still exhibits an exponential decrease, with a finite quantum coherence length $\xi_Q$ diverging only upon reaching $T=0$ \cite{MalpettiR2016}. This exhibits a most dramatic separation of scale between total and quantum correlations, and suggests again that a semiclassical approach truncating quantum correlations has the potential to quantitatively describe the correlations in this system.  
 
 Fig.~\ref{f.corrQMF}(a-c) shows how the description of correlations within the cQMF approximation evolve with the cluster size in the case of a strongly interacting, yet superfluid regime ($u=3$). It is clear that, as seen already for quantum Ising spins, the cQMF approximation overestimates the strength of correlations by underestimating quantum effects, and that, for most values of the distance $r$ (namely for $r \geq 2$), the correlation function $g(r)$ obtained via cQMF converges from above to the exact result. Yet, plotting the correlation functions in log-log scale shows something rather remarkable: despite the clear difference in the very value of the correlation function, the long-range tails of the correlation functions for different $l_c$ values appear as parallel in log-log scale, suggesting that the asymptotic behavior of the correlation function is captured by the cQMF up to a multiplicative constant, which accounts for the incomplete description of short-range quantum fluctuations. The ability of the cQMF to capture quantitatively the long-range correlations -- whose decay comes entirely from thermal fluctuations -- is best seen by fitting the correlation function to a power-law form with periodic boundary conditions, $A/d(r|L)^{\eta}$ where $d(r|L) = (L/\pi) \sin(\pi r/L)$ is the chord length. The so-extracted $\eta$ exponent is shown in Fig.~\ref{f.corrQMF}(d) as a function of temperature: we observe that using the cQMF for the smallest cluster $l_c=1$ already produces a very accurate result compared to the exact one. This is by no means a trivial result, as quantum fluctuations are indeed strong in this example, speeding up significantly the increase of the $\eta$ exponent with respect to the classical limit (shown in the same figure as the spin-wave result $\eta= t/(2\pi)$, valid at low $T$ for $u=0$). This provides an interesting insight into the physics of the model at hand, showing that, sufficiently far from the quantum critical point $u=u_c$, the quantum renormalization of the $\eta$ exponent for the power-law decay of superfluid correlations comes from very-short-ranged quantum fluctuations: such fluctuations are indeed captured faithfully by an approximation that discards any quantum correlation between different sites (namely the cQMF approach with $l_c=1$). 
 
 The above picture suggests that the long-distance behavior of 2$d$ quantum rotors with $u=3$ (and lower) is essentially captured by an effective classical model of time-averaged rotors 
 \begin{equation}
 S_{\rm eff} = - \frac{K}{2} \sum_{\langle ij \rangle}  \left( \overline{e^{i\theta_i}} ~~\overline{e^{-i\theta_j}} + c.c. \right ) - K_{\tau} \sum_{i,k} \cos(\theta_{i,k} - \theta_{i,k+1}) 
  \end{equation}
  \label{e.SeffQR}
  similarly to what seen for quantum Ising spins in a moderate transverse field (see Eq.~\eqref{e.SeffIsing}). By definition, the correlations among the time-averaged degrees of freedom are captured by the thermal part ($g_T$) of the correlation function (Eq.~\eqref{e.classcorr}). Hence the success of cQMF with $l_c=1$ implies that the classical correlations $g_T$ of the full quantum model should be accurately described by the cQMF approximation, but only up to an overall multiplicative factor which accounts for the fact that, in the real system, the effective classical degrees of freedom are recovered by tracing out quantum fluctuations correlated within a volume $\sim \xi_Q^2$.    
 
 To test this implication, we investigate the normalized thermal correlation function $g_T(r)/g_T(L/2)$, where the normalization to the asymptotic value eliminates the above-mentioned multiplicative factor. Fig.~\ref{f.corr_anatomy1}(a) compares the normalized thermal correlations for different cluster sizes with the exact result: remarkably the \emph{full} structure of the thermal correlation function is well captured by the cQMF approach. Moreover it is striking to observe that the thermal correlation function exhibits a power-law behavior (very well fitted to $A/d(r|L)^{\eta}$) across the entire range of separations $r$, whereas the total correlation function clearly deviates from a simple power law, as shown in Fig.~\ref{f.corrQMF}.  This implies that the short-range deviation from a power-law entirely comes from the quantum correlations, which are progressively captured by the cQMF approach with an increasingly large cluster size (Fig.~\ref{f.corr_anatomy1}(b)). Fig.~\ref{f.corr_anatomy2} finally puts all the pieces together, showing that the total correlation function for critical superfluid correlations of 2$d$ quantum rotors reproduces the power-law thermal correlation function at large distances, once an exponentially decaying short-range quantum component has died out over a distance $r\gg \xi_Q$ for  the real system, or a distance $r \geq l_c$ for its cQMF approximation, in agreement with the cartoon proposed by Fig.~\ref{f.corr}.  
 
 \begin{figure}[ht!]
\includegraphics[width = \linewidth]{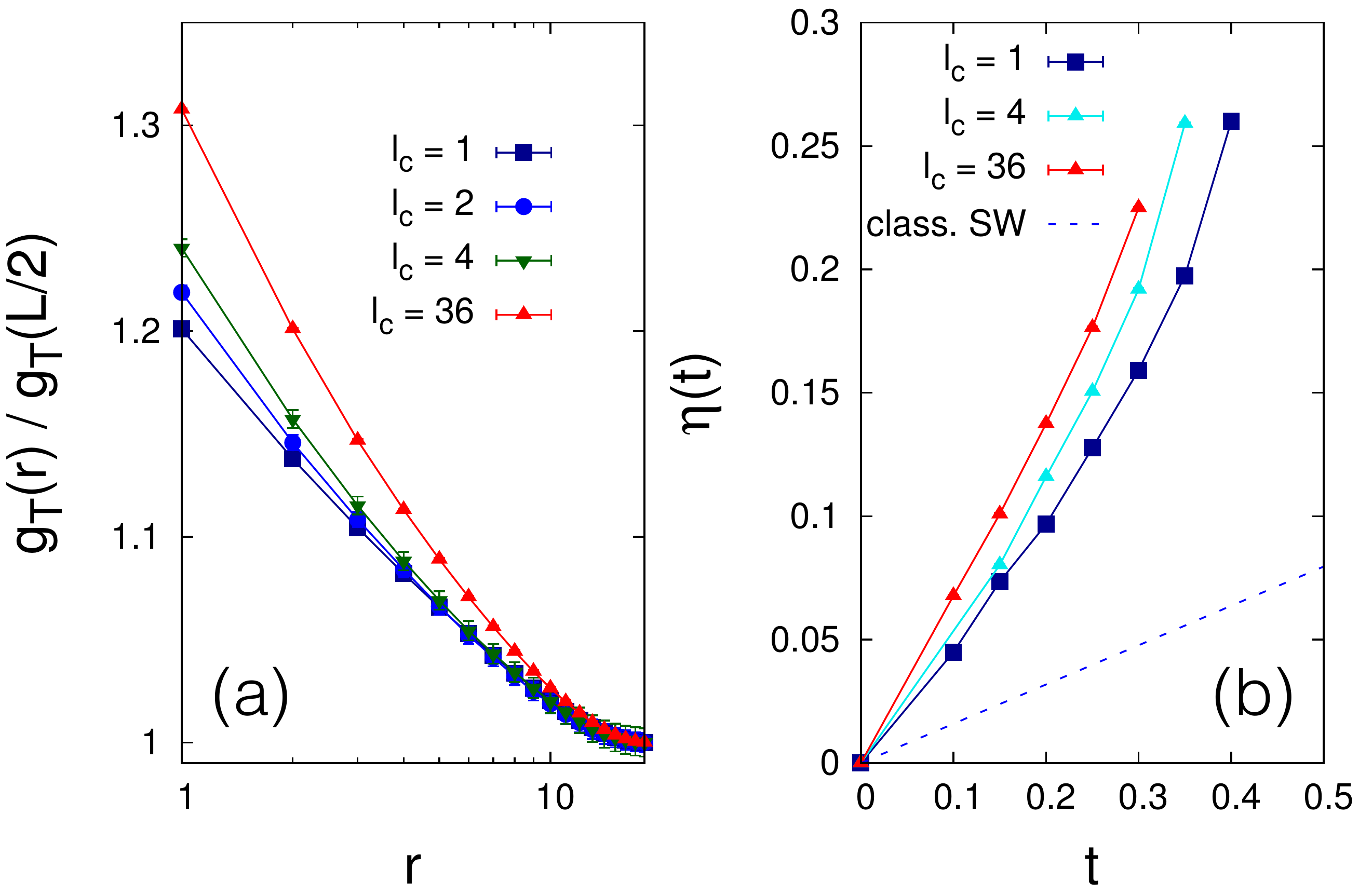}
\caption{Correlations in the superfluid phase of 2$d$ quantum rotors with $u=5$, $L=36$. (a) Normalized thermal correlations at $t=0.15$ for various cluster sizes. (b) $\eta$ exponent extracted from power-law fits of the thermal correlation function over the range $[3,L/2]$.}
\label{f.corr_u5}
\end{figure}
 
  The ability of the cQMF approximation to accurately capture the thermal correlations up to a global prefactor -- and hence the total correlations at long distance -- is altered progressively in the low-temperature superfluid phase as one approaches the quantum critical point $u = u_c$.  Indeed, in that limit $\xi_Q$ grows progressively and the effective classical degrees of freedom, whose correlations are probed by $g(r)$ (or $g_T(r)$) at $r\gg \xi_Q$, emerge from integrating out short-range quantum fluctuations over increasingly large quantum-correlated volumes. This means that the exponent $\eta$ for the decay of $g(r)$ is affected by increasingly longer-range quantum fluctuations, and a simple cQMF approach with $l_c=1$ may fail. This is shown in Fig.~\ref{f.corr_u5} for the case $u=5$ (approaching $u_c = 5.8$) where one can see that, in order to capture the correct $\eta$ exponent, it is necessary to take into account quantum fluctuations on a sizable correlation volume, namely using $l_c>1$, and that, as already observed in the case of quantum Ising spins close to their quantum critical point, convergence in the cluster size is rather slow.    
 
 \begin{figure}[ht!]
\includegraphics[width = \linewidth]{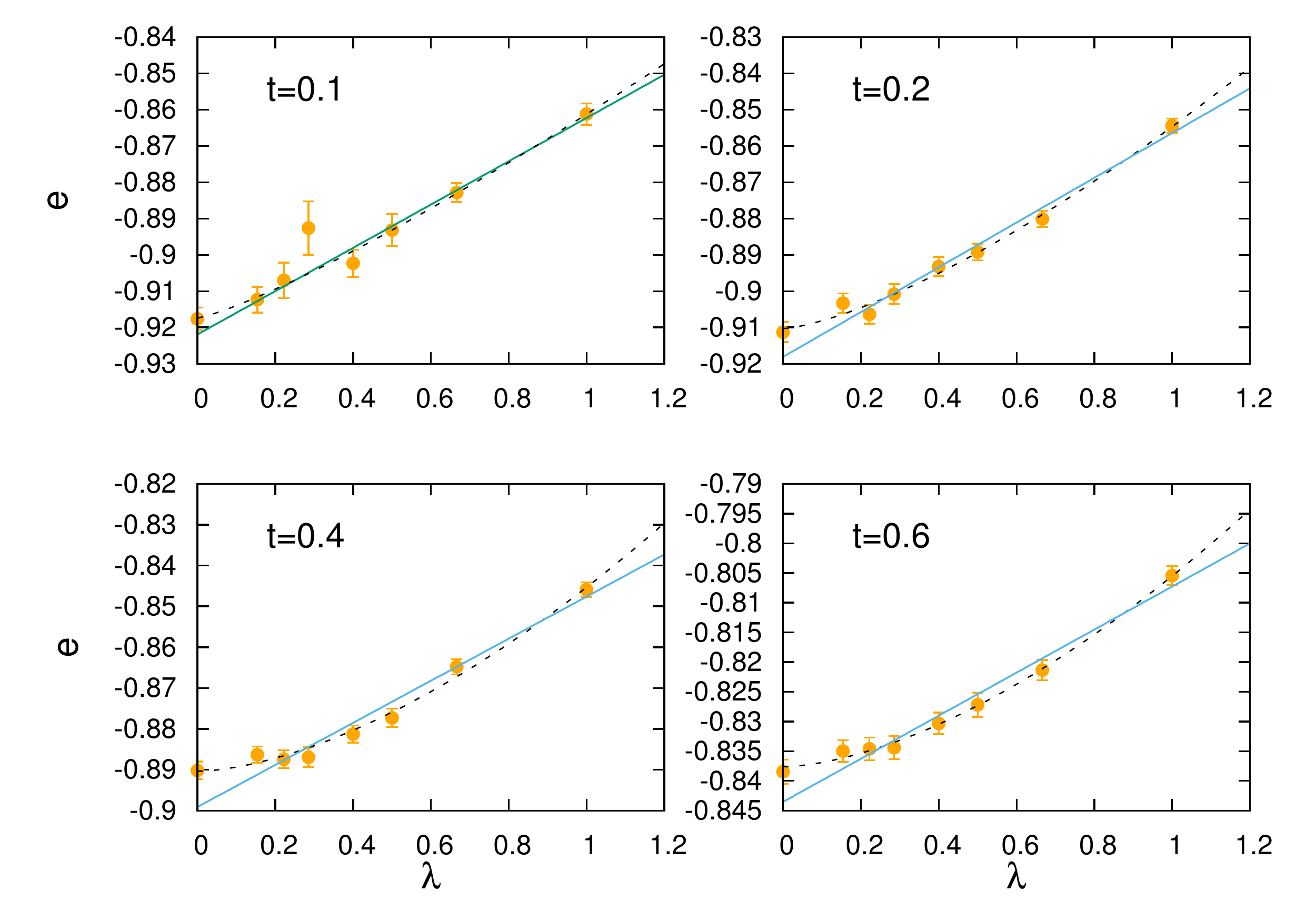}
\caption{Scaling of the energy density $e$ with the cluster parameter $\lambda$ (see text) for quantum rotors on the square lattice. $\lambda$ values correspond to cluster linear sizes $l_c =$ 1, 2, 3, 4, 6, 12 and 24 (in decreasing order).
The solid and dashed lines are a linear fit (excluding the points at $\lambda = 0$ and 1) and a power-law fit $a_1+a_2*\lambda^{a_3}$ (retaining all points), respectively.  Model parameters are $u=3$ and $L=24$.}
\label{f.enscaling}
\end{figure}

\begin{figure}[ht!]
\includegraphics[width = \linewidth]{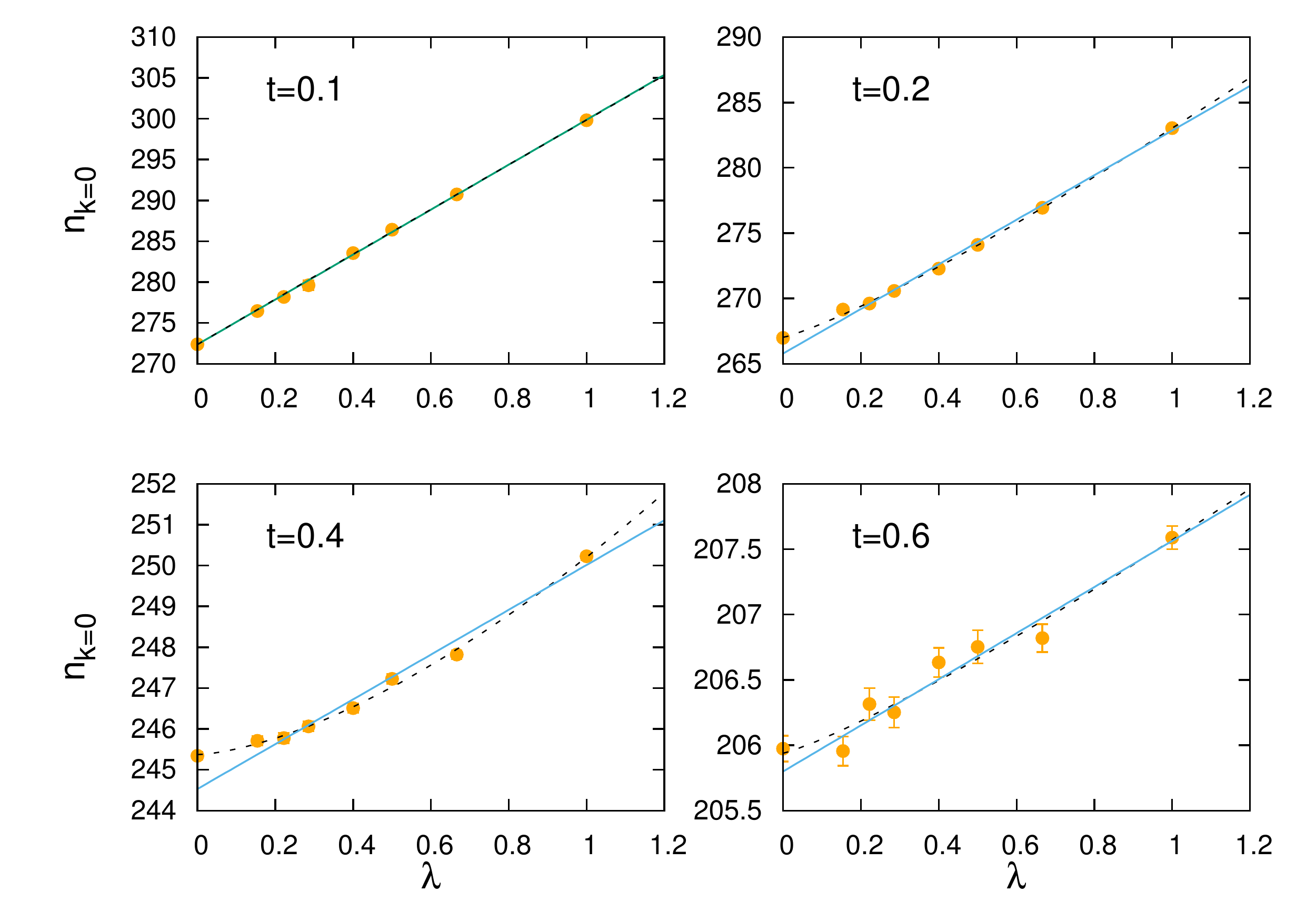}
\caption{Scaling of the condensate peak $n_{k=0}$ with the cluster parameter $\lambda$ for quantum rotors on the square lattice. Lines and parameters as in Fig.~\ref{f.enscaling}.}
\label{f.nk0scaling}
\end{figure}

\begin{figure}[ht!]
\includegraphics[width = \linewidth]{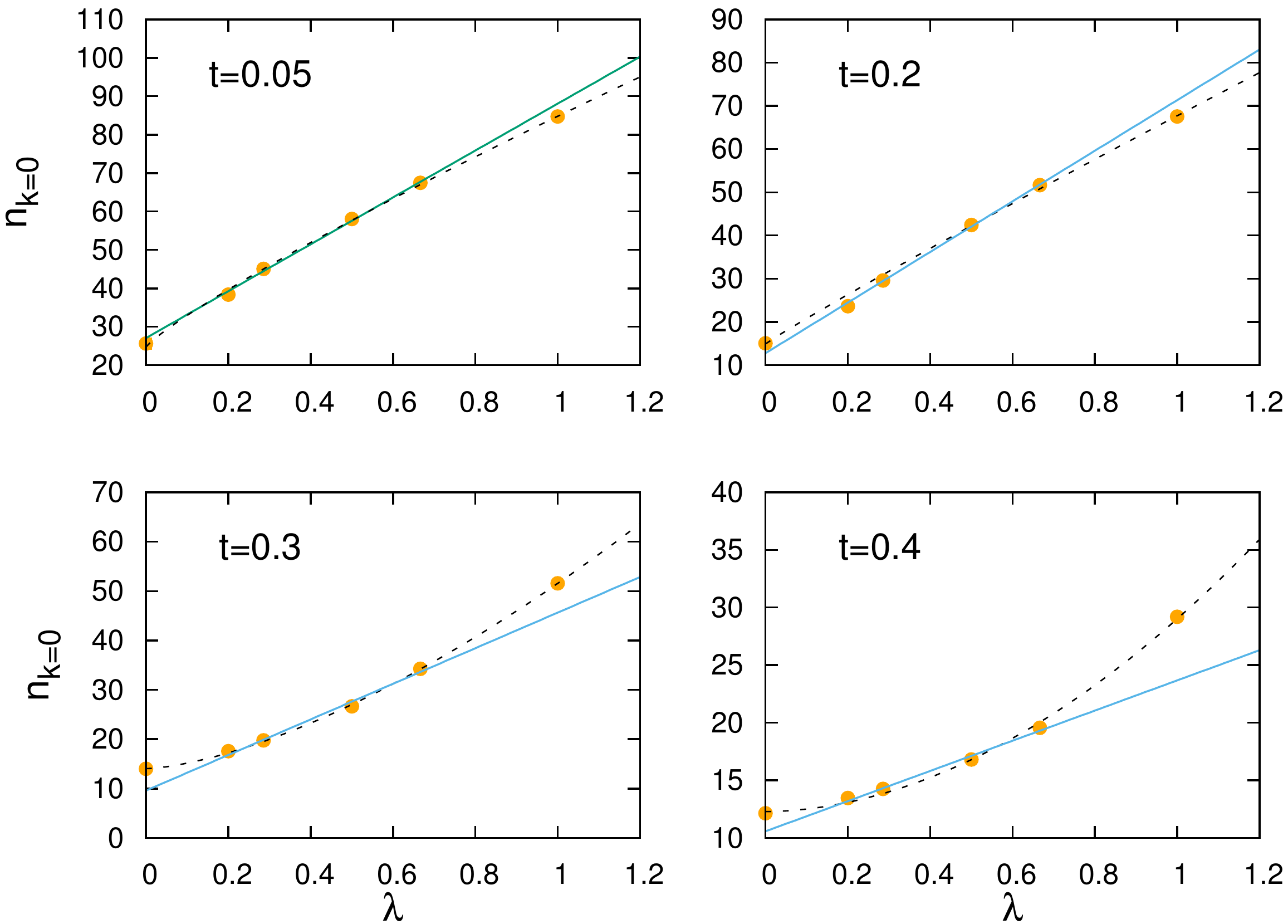}
\caption{Scaling of the condensate peak $n_{k=0}$ with the cluster parameter $\lambda$ for quantum rotors on the square lattice. $\lambda$ values correspond to cluster linear sizes $l_c =$ 1, 2, 3, 6, 9, and 18 (in decreasing order).
Model parameters are $u=5.8$ and $L=18$.
 Lines as in Fig.~\ref{f.nk0scaling}.}
\label{f.nk0scalingU5.8}
\end{figure}

 \subsubsection{Cluster-size scaling}
  To determine the precise form of the convergence of observables upon increasing the cluster size, we focus our attention on the energy density $e = \langle {\cal H} \rangle / L^2$ and the $k=0$ peak in the momentum distribution
 \begin{equation}
  n_{k=0} =  \frac{1}{L^2} \sum_{ij}  \langle e^{i(\theta_i - \theta_j)} \rangle~.
  \end{equation}
  Figs.~\ref{f.enscaling} and \ref{f.nk0scaling} show the cluster-size scaling of the above quantities in the superfluid phase of 2$d$ quantum rotors for $u=3$. Similarly to the case of quantum Ising spins, we observe that, when plotted against the surface-to-bulk ratio $\lambda$, the above mentioned quantities scale generically as a power law towards the exact result for $\lambda = 0$: this again validates the choice of the $\lambda$ parameter to extrapolate the finite-cluster results towards the exact one. In particular the exponent of the power law seemingly approaches one when $T\to 0$: this is consistent with the numerical observation of linear scaling in ground-state studies of lattice boson models, based on the cMFT \cite{Luehmann2013,Yamamotoetal2013, Trousseletetal2014, Yamamotoetal2015}. On the other hand, the exponent appears to grow gradually upon increasing the temperature, namely the convergence towards the exact result is slightly faster, the higher the temperature.  
  
  Finally it is interesting to observe that the above-cited cluster-size scaling of the cQMF results holds not only in the superfluid phase, but also along the quantum-critical trajectory, namely for finite temperatures above the quantum critical point $u = u_c \approx 5.8$ for the superfluid/Mott-insulator transition. As shown in Fig.~\ref{f.nk0scalingU5.8}, a power-law scaling is well consistent with the numerical results, with a nearly linear scaling which persists at higher temperatures, and a much larger prefactor. The slower scaling reveals a strong enhancement of quantum correlations in the quantum critical region, in spite of the fact that the total correlations have in fact acquired an exponentially decreasing form at finite temperature.
 
 \section{Conclusions}
 \label{s.conclusions}
 
 In this paper we have developed a path-integral picture of the recently proposed \cite{MalpettiR2016} separation between classical/thermal and quantum correlations in equilibrium quantum many-body systems. Based on that picture, a new approximation scheme -- called the (cluster) quantum mean-field (cQMF) approximation -- can be introduced, which acts as a mean-field approximation \emph{restricted} to quantum correlations, discarding them beyond the length of the clusters it uses. The rationale of such an approximation relies on the fact that quantum correlations are short-ranged at any finite temperature -- exhibiting exponential decay over a finite quantum coherence length -- whereas classical/thermal correlations can have an arbitrarily long range. The cQMF is then able to describe thermal correlations faithfully, while partially account for the renormalization effects coming from short-range quantum fluctuations. We have developed this approximation in the case of two paradigmatic lattice quantum models, namely the transverse-field Ising model and the quantum-rotor model on the square lattice, and exhibited the insight that the cQMF provides in the structure of correlations. 
  
   Throughout this paper the cQMF approach was implemented within a path-integral Monte Carlo approach which is otherwise able to provide (numerically) exact results for the equilibrium properties of the systems under investigation. Therefore the cQMF approximation was used as a ``ruler" in order to measure the impact of quantum correlations on the thermodynamics of strongly correlated quantum systems: we systematically compared the exact results with the ``caricature" offered by cQMF of quantum correlations restricted to spatially separated clusters. In so doing we showed that, sufficiently far from quantum critical points, the cQMF approximation can be extremely accurate, revealing that short-range quantum fluctuations restricted to very small clusters already offer a quantitative account of the actual quantum fluctuations in the real system. At the same time, the insight gained with our study paves the way to the development of semiclassical approaches exploiting the separation of scale between classical and quantum correlations; and implementing the cQMF approximation in a way which allows one to tackle quantum many-body models which are otherwise untreatable. A first example of such an approach will be presented in a forthcoming publication \cite{AFMC}.

\section{Acknowledgements}  We thank A. Ran\c con and I. Fr\'erot for useful discussions and collaboration on related projects. This work is supported by ANR (``ArtiQ" project).  All simulations were performed on the PSMN cluster of the ENS of Lyon.

\bibliography{AFMC,qv}

\begin{thebibliography}{32}
\expandafter\ifx\csname natexlab\endcsname\relax\def\natexlab#1{#1}\fi
\expandafter\ifx\csname bibnamefont\endcsname\relax
  \def\bibnamefont#1{#1}\fi
\expandafter\ifx\csname bibfnamefont\endcsname\relax
  \def\bibfnamefont#1{#1}\fi
\expandafter\ifx\csname citenamefont\endcsname\relax
  \def\citenamefont#1{#1}\fi
\expandafter\ifx\csname url\endcsname\relax
  \def\url#1{\texttt{#1}}\fi
\expandafter\ifx\csname urlprefix\endcsname\relax\def\urlprefix{URL }\fi
\providecommand{\bibinfo}[2]{#2}
\providecommand{\eprint}[2][]{\url{#2}}

\bibitem[{\citenamefont{Goldenfeld}(1992)}]{Goldenfeldbook}
\bibinfo{author}{\bibfnamefont{N.}~\bibnamefont{Goldenfeld}},
  \emph{\bibinfo{title}{Lectures On Phase Transitions And The Renormalization
  Group}} (\bibinfo{publisher}{Addison-Wesley}, \bibinfo{year}{1992}).

\bibitem[{\citenamefont{Fisher et~al.}(1989)\citenamefont{Fisher, Weichman,
  Grinstein, and Fisher}}]{Fisheretal1989}
\bibinfo{author}{\bibfnamefont{M.~P.~A.} \bibnamefont{Fisher}},
  \bibinfo{author}{\bibfnamefont{P.~B.} \bibnamefont{Weichman}},
  \bibinfo{author}{\bibfnamefont{G.}~\bibnamefont{Grinstein}},
  \bibnamefont{and} \bibinfo{author}{\bibfnamefont{D.~S.}
  \bibnamefont{Fisher}}, \bibinfo{journal}{Phys. Rev. B}
  \textbf{\bibinfo{volume}{40}}, \bibinfo{pages}{546} (\bibinfo{year}{1989}),
  \urlprefix\url{http://link.aps.org/doi/10.1103/PhysRevB.40.546}.

\bibitem[{\citenamefont{Rokhsar and Kotliar}(1991)}]{RokhsarK1991}
\bibinfo{author}{\bibfnamefont{D.~S.} \bibnamefont{Rokhsar}} \bibnamefont{and}
  \bibinfo{author}{\bibfnamefont{B.~G.} \bibnamefont{Kotliar}},
  \bibinfo{journal}{Phys. Rev. B} \textbf{\bibinfo{volume}{44}},
  \bibinfo{pages}{10328} (\bibinfo{year}{1991}),
  \urlprefix\url{http://link.aps.org/doi/10.1103/PhysRevB.44.10328}.

\bibitem[{\citenamefont{Bethe}(1935)}]{Bethe1935}
\bibinfo{author}{\bibfnamefont{H.~A.} \bibnamefont{Bethe}},
  \bibinfo{journal}{Proceedings of the Royal Society of London A: Mathematical,
  Physical and Engineering Sciences} \textbf{\bibinfo{volume}{150}},
  \bibinfo{pages}{552} (\bibinfo{year}{1935}), ISSN \bibinfo{issn}{0080-4630},
  \urlprefix\url{http://rspa.royalsocietypublishing.org/content/150/871/552}.

\bibitem[{\citenamefont{Peierls}(1936)}]{Peierls1936}
\bibinfo{author}{\bibfnamefont{R.}~\bibnamefont{Peierls}},
  \bibinfo{journal}{Mathematical Proceedings of the Cambridge Philosophical
  Society} \textbf{\bibinfo{volume}{32}}, \bibinfo{pages}{477}
  (\bibinfo{year}{1936}), ISSN \bibinfo{issn}{1469-8064},
  \urlprefix\url{http://journals.cambridge.org/article_S0305004100019174}.

\bibitem[{\citenamefont{Newman and Barkema}(1999)}]{NewmanBbook}
\bibinfo{author}{\bibfnamefont{M.~E.~J.} \bibnamefont{Newman}}
  \bibnamefont{and} \bibinfo{author}{\bibfnamefont{G.~T.}
  \bibnamefont{Barkema}}, \emph{\bibinfo{title}{Monte Carlo Methods in
  Statistical Physics}} (\bibinfo{publisher}{Oxford}, \bibinfo{year}{1999}).

\bibitem[{\citenamefont{Chandrasekharan and
  Wiese}(1999)}]{ChandrasekharanW1999}
\bibinfo{author}{\bibfnamefont{S.}~\bibnamefont{Chandrasekharan}}
  \bibnamefont{and} \bibinfo{author}{\bibfnamefont{U.-J.} \bibnamefont{Wiese}},
  \bibinfo{journal}{Phys. Rev. Lett.} \textbf{\bibinfo{volume}{83}},
  \bibinfo{pages}{3116} (\bibinfo{year}{1999}),
  \urlprefix\url{http://link.aps.org/doi/10.1103/PhysRevLett.83.3116}.

\bibitem[{\citenamefont{Henelius and Sandvik}(2000)}]{HeneliusS2000}
\bibinfo{author}{\bibfnamefont{P.}~\bibnamefont{Henelius}} \bibnamefont{and}
  \bibinfo{author}{\bibfnamefont{A.~W.} \bibnamefont{Sandvik}},
  \bibinfo{journal}{Phys. Rev. B} \textbf{\bibinfo{volume}{62}},
  \bibinfo{pages}{1102} (\bibinfo{year}{2000}),
  \urlprefix\url{http://link.aps.org/doi/10.1103/PhysRevB.62.1102}.

\bibitem[{\citenamefont{{Malpetti} and
  {Roscilde}}(2016{\natexlab{a}})}]{MalpettiR2016}
\bibinfo{author}{\bibfnamefont{D.}~\bibnamefont{{Malpetti}}} \bibnamefont{and}
  \bibinfo{author}{\bibfnamefont{T.}~\bibnamefont{{Roscilde}}},
  \bibinfo{journal}{ArXiv e-prints}  (\bibinfo{year}{2016}{\natexlab{a}}),
  \eprint{1605.04223}.

\bibitem[{\citenamefont{{Malpetti} and {Roscilde}}(2016{\natexlab{b}})}]{AFMC}
\bibinfo{author}{\bibfnamefont{D.}~\bibnamefont{{Malpetti}}} \bibnamefont{and}
  \bibinfo{author}{\bibfnamefont{T.}~\bibnamefont{{Roscilde}}},
  \bibinfo{journal}{in preparation}  (\bibinfo{year}{2016}{\natexlab{b}}).

\bibitem[{\citenamefont{Cuccoli et~al.}(1995)\citenamefont{Cuccoli, Giachetti,
  Tognetti, Vaia, and Verrucchi}}]{Cuccolietal1995}
\bibinfo{author}{\bibfnamefont{A.}~\bibnamefont{Cuccoli}},
  \bibinfo{author}{\bibfnamefont{R.}~\bibnamefont{Giachetti}},
  \bibinfo{author}{\bibfnamefont{V.}~\bibnamefont{Tognetti}},
  \bibinfo{author}{\bibfnamefont{R.}~\bibnamefont{Vaia}}, \bibnamefont{and}
  \bibinfo{author}{\bibfnamefont{P.}~\bibnamefont{Verrucchi}},
  \bibinfo{journal}{Journal of Physics: Condensed Matter}
  \textbf{\bibinfo{volume}{7}}, \bibinfo{pages}{7891} (\bibinfo{year}{1995}),
  \urlprefix\url{http://stacks.iop.org/0953-8984/7/i=41/a=003}.

\bibitem[{\citenamefont{Feynman and Kleinert}(1986)}]{FeynmanK1986}
\bibinfo{author}{\bibfnamefont{R.~P.} \bibnamefont{Feynman}} \bibnamefont{and}
  \bibinfo{author}{\bibfnamefont{H.}~\bibnamefont{Kleinert}},
  \bibinfo{journal}{Phys. Rev. A} \textbf{\bibinfo{volume}{34}},
  \bibinfo{pages}{5080} (\bibinfo{year}{1986}),
  \urlprefix\url{http://link.aps.org/doi/10.1103/PhysRevA.34.5080}.

\bibitem[{\citenamefont{Giuliani and Vignale}(2008)}]{Vignalebook}
\bibinfo{author}{\bibfnamefont{G.}~\bibnamefont{Giuliani}} \bibnamefont{and}
  \bibinfo{author}{\bibfnamefont{G.}~\bibnamefont{Vignale}},
  \emph{\bibinfo{title}{Quantum Theory of the Electron Liquid}}
  (\bibinfo{publisher}{Cambridge}, \bibinfo{year}{2008}).

\bibitem[{\citenamefont{Horodecki et~al.}(2009)\citenamefont{Horodecki,
  Horodecki, Horodecki, and Horodecki}}]{Horodecki2009}
\bibinfo{author}{\bibfnamefont{R.}~\bibnamefont{Horodecki}},
  \bibinfo{author}{\bibfnamefont{P.}~\bibnamefont{Horodecki}},
  \bibinfo{author}{\bibfnamefont{M.}~\bibnamefont{Horodecki}},
  \bibnamefont{and}
  \bibinfo{author}{\bibfnamefont{K.}~\bibnamefont{Horodecki}},
  \bibinfo{journal}{Rev. Mod. Phys.} \textbf{\bibinfo{volume}{81}},
  \bibinfo{pages}{865} (\bibinfo{year}{2009}),
  \urlprefix\url{http://link.aps.org/doi/10.1103/RevModPhys.81.865}.

\bibitem[{\citenamefont{McIntosh et~al.}(2012)\citenamefont{McIntosh, Pisarski,
  Gooding, and Zaremba}}]{McIntoshetal2012}
\bibinfo{author}{\bibfnamefont{T.}~\bibnamefont{McIntosh}},
  \bibinfo{author}{\bibfnamefont{P.}~\bibnamefont{Pisarski}},
  \bibinfo{author}{\bibfnamefont{R.~J.} \bibnamefont{Gooding}},
  \bibnamefont{and} \bibinfo{author}{\bibfnamefont{E.}~\bibnamefont{Zaremba}},
  \bibinfo{journal}{Phys. Rev. A} \textbf{\bibinfo{volume}{86}},
  \bibinfo{pages}{013623} (\bibinfo{year}{2012}),
  \urlprefix\url{http://link.aps.org/doi/10.1103/PhysRevA.86.013623}.

\bibitem[{\citenamefont{L\"uhmann}(2013)}]{Luehmann2013}
\bibinfo{author}{\bibfnamefont{D.-S.} \bibnamefont{L\"uhmann}},
  \bibinfo{journal}{Phys. Rev. A} \textbf{\bibinfo{volume}{87}},
  \bibinfo{pages}{043619} (\bibinfo{year}{2013}),
  \urlprefix\url{http://link.aps.org/doi/10.1103/PhysRevA.87.043619}.

\bibitem[{\citenamefont{Yamamoto et~al.}(2014)\citenamefont{Yamamoto,
  Marmorini, and Danshita}}]{Yamamotoetal2013}
\bibinfo{author}{\bibfnamefont{D.}~\bibnamefont{Yamamoto}},
  \bibinfo{author}{\bibfnamefont{G.}~\bibnamefont{Marmorini}},
  \bibnamefont{and} \bibinfo{author}{\bibfnamefont{I.}~\bibnamefont{Danshita}},
  \bibinfo{journal}{Phys. Rev. Lett.} \textbf{\bibinfo{volume}{112}},
  \bibinfo{pages}{127203} (\bibinfo{year}{2014}),
  \urlprefix\url{http://link.aps.org/doi/10.1103/PhysRevLett.112.127203}.

\bibitem[{\citenamefont{Trousselet et~al.}(2014)\citenamefont{Trousselet,
  Rueda-Fonseca, and Ralko}}]{Trousseletetal2014}
\bibinfo{author}{\bibfnamefont{F.}~\bibnamefont{Trousselet}},
  \bibinfo{author}{\bibfnamefont{P.}~\bibnamefont{Rueda-Fonseca}},
  \bibnamefont{and} \bibinfo{author}{\bibfnamefont{A.}~\bibnamefont{Ralko}},
  \bibinfo{journal}{Phys. Rev. B} \textbf{\bibinfo{volume}{89}},
  \bibinfo{pages}{085104} (\bibinfo{year}{2014}),
  \urlprefix\url{http://link.aps.org/doi/10.1103/PhysRevB.89.085104}.

\bibitem[{\citenamefont{Yamamoto et~al.}(2015)\citenamefont{Yamamoto,
  Marmorini, and Danshita}}]{Yamamotoetal2015}
\bibinfo{author}{\bibfnamefont{D.}~\bibnamefont{Yamamoto}},
  \bibinfo{author}{\bibfnamefont{G.}~\bibnamefont{Marmorini}},
  \bibnamefont{and} \bibinfo{author}{\bibfnamefont{I.}~\bibnamefont{Danshita}},
  \bibinfo{journal}{Phys. Rev. Lett.} \textbf{\bibinfo{volume}{114}},
  \bibinfo{pages}{027201} (\bibinfo{year}{2015}),
  \urlprefix\url{http://link.aps.org/doi/10.1103/PhysRevLett.114.027201}.

\bibitem[{\citenamefont{Negele and Orland}(1998)}]{NegeleO1988}
\bibinfo{author}{\bibfnamefont{J.~W.} \bibnamefont{Negele}} \bibnamefont{and}
  \bibinfo{author}{\bibfnamefont{H.}~\bibnamefont{Orland}},
  \emph{\bibinfo{title}{Quantum Many-Particle Systems}}
  (\bibinfo{publisher}{Perseus Books}, \bibinfo{year}{1998}).

\bibitem[{\citenamefont{Pitaevskii and Stringari}(2016)}]{StringariPbook}
\bibinfo{author}{\bibfnamefont{L.}~\bibnamefont{Pitaevskii}} \bibnamefont{and}
  \bibinfo{author}{\bibfnamefont{S.}~\bibnamefont{Stringari}},
  \emph{\bibinfo{title}{Bose-Einstein Condensation and Superfluidity}}
  (\bibinfo{publisher}{Oxford}, \bibinfo{year}{2016}).

\bibitem[{\citenamefont{Castin}(2004)}]{Castin2004}
\bibinfo{author}{\bibfnamefont{Y.}~\bibnamefont{Castin}},
  \bibinfo{journal}{{Journal de Physique IV Colloque}}
  \textbf{\bibinfo{volume}{116}}, \bibinfo{pages}{89} (\bibinfo{year}{2004}),
  \bibinfo{note}{lecture notes of Les Houches school on low dimensional quantum
  gases (April 2003), M. Olshanii, H. Perrin, L. Pricoupenko, Eds.},
  \urlprefix\url{https://hal.archives-ouvertes.fr/hal-00002178}.

\bibitem[{\citenamefont{Blakie€  et~al.}(2008)\citenamefont{Blakie€ ,
  Bradley€ , Davis, Ballagh, and Gardiner}}]{Blakieetal2008}
\bibinfo{author}{\bibfnamefont{P.}~\bibnamefont{Blakie€ }},
  \bibinfo{author}{\bibfnamefont{A.}~\bibnamefont{Bradley€ }},
  \bibinfo{author}{\bibfnamefont{M.}~\bibnamefont{Davis}},
  \bibinfo{author}{\bibfnamefont{R.}~\bibnamefont{Ballagh}}, \bibnamefont{and}
  \bibinfo{author}{\bibfnamefont{C.}~\bibnamefont{Gardiner}},
  \bibinfo{journal}{Advances in Physics} \textbf{\bibinfo{volume}{57}},
  \bibinfo{pages}{363} (\bibinfo{year}{2008}),
  \eprint{http://dx.doi.org/10.1080/00018730802564254},
  \urlprefix\url{http://dx.doi.org/10.1080/00018730802564254}.

\bibitem[{\citenamefont{Suzuki}(1976)}]{Suzuki1976}
\bibinfo{author}{\bibfnamefont{M.}~\bibnamefont{Suzuki}},
  \bibinfo{journal}{Prog. Theor. Phys.} \textbf{\bibinfo{volume}{56}},
  \bibinfo{pages}{1454} (\bibinfo{year}{1976}).

\bibitem[{\citenamefont{Werner}(1989)}]{Werner1989}
\bibinfo{author}{\bibfnamefont{R.~F.} \bibnamefont{Werner}},
  \bibinfo{journal}{Phys. Rev. A} \textbf{\bibinfo{volume}{40}},
  \bibinfo{pages}{4277} (\bibinfo{year}{1989}),
  \urlprefix\url{http://link.aps.org/doi/10.1103/PhysRevA.40.4277}.

\bibitem[{\citenamefont{Feynman et~al.}(2005)\citenamefont{Feynman, Hibbs, and
  Styer}}]{FeynmanHibbs}
\bibinfo{author}{\bibfnamefont{R.~P.} \bibnamefont{Feynman}},
  \bibinfo{author}{\bibfnamefont{A.~R.} \bibnamefont{Hibbs}}, \bibnamefont{and}
  \bibinfo{author}{\bibfnamefont{D.~F.} \bibnamefont{Styer}},
  \emph{\bibinfo{title}{Quantum Mechanics and Path Integrals}}
  (\bibinfo{publisher}{Dover}, \bibinfo{year}{2005}).

\bibitem[{\citenamefont{Dutta et~al.}(2015)\citenamefont{Dutta, Aeppli,
  Chakrabarti, Divakaran, Rosenbaum, and Sen}}]{quantumIsingbook}
\bibinfo{author}{\bibfnamefont{A.}~\bibnamefont{Dutta}},
  \bibinfo{author}{\bibfnamefont{G.}~\bibnamefont{Aeppli}},
  \bibinfo{author}{\bibfnamefont{B.~K.} \bibnamefont{Chakrabarti}},
  \bibinfo{author}{\bibfnamefont{U.}~\bibnamefont{Divakaran}},
  \bibinfo{author}{\bibfnamefont{T.~F.} \bibnamefont{Rosenbaum}},
  \bibnamefont{and} \bibinfo{author}{\bibfnamefont{D.}~\bibnamefont{Sen}},
  \emph{\bibinfo{title}{Quantum Phase Transitions in Transverse Field Spin
  Models}} (\bibinfo{publisher}{Cambridge}, \bibinfo{year}{2015}).

\bibitem[{\citenamefont{Anderson}(1952)}]{Anderson1952}
\bibinfo{author}{\bibfnamefont{P.~W.} \bibnamefont{Anderson}},
  \bibinfo{journal}{Phys. Rev.} \textbf{\bibinfo{volume}{86}},
  \bibinfo{pages}{694} (\bibinfo{year}{1952}),
  \urlprefix\url{http://link.aps.org/doi/10.1103/PhysRev.86.694}.

\bibitem[{\citenamefont{Bl\"ote and Deng}(2002)}]{DengB2002}
\bibinfo{author}{\bibfnamefont{H.~W.~J.} \bibnamefont{Bl\"ote}}
  \bibnamefont{and} \bibinfo{author}{\bibfnamefont{Y.}~\bibnamefont{Deng}},
  \bibinfo{journal}{Phys. Rev. E} \textbf{\bibinfo{volume}{66}},
  \bibinfo{pages}{066110} (\bibinfo{year}{2002}),
  \urlprefix\url{http://link.aps.org/doi/10.1103/PhysRevE.66.066110}.

\bibitem[{\citenamefont{Wolff}(1989)}]{Wolff1989}
\bibinfo{author}{\bibfnamefont{U.}~\bibnamefont{Wolff}},
  \bibinfo{journal}{Phys. Rev. Lett.} \textbf{\bibinfo{volume}{62}},
  \bibinfo{pages}{361} (\bibinfo{year}{1989}),
  \urlprefix\url{http://link.aps.org/doi/10.1103/PhysRevLett.62.361}.

\bibitem[{\citenamefont{Wallin et~al.}(1994)\citenamefont{Wallin, S{\o}rensen,
  Girvin, and Young}}]{Wallinetal1994}
\bibinfo{author}{\bibfnamefont{M.}~\bibnamefont{Wallin}},
  \bibinfo{author}{\bibfnamefont{E.~S.} \bibnamefont{S{\o}rensen}},
  \bibinfo{author}{\bibfnamefont{S.~M.} \bibnamefont{Girvin}},
  \bibnamefont{and} \bibinfo{author}{\bibfnamefont{A.~P.} \bibnamefont{Young}},
  \bibinfo{journal}{Phys. Rev. B} \textbf{\bibinfo{volume}{49}},
  \bibinfo{pages}{12115} (\bibinfo{year}{1994}),
  \urlprefix\url{http://link.aps.org/doi/10.1103/PhysRevB.49.12115}.

\bibitem[{\citenamefont{Sondhi et~al.}(1997)\citenamefont{Sondhi, Girvin,
  Carini, and Shahar}}]{Sondhietal1997}
\bibinfo{author}{\bibfnamefont{S.~L.} \bibnamefont{Sondhi}},
  \bibinfo{author}{\bibfnamefont{S.~M.} \bibnamefont{Girvin}},
  \bibinfo{author}{\bibfnamefont{J.~P.} \bibnamefont{Carini}},
  \bibnamefont{and} \bibinfo{author}{\bibfnamefont{D.}~\bibnamefont{Shahar}},
  \bibinfo{journal}{Rev. Mod. Phys.} \textbf{\bibinfo{volume}{69}},
  \bibinfo{pages}{315} (\bibinfo{year}{1997}),
  \urlprefix\url{http://link.aps.org/doi/10.1103/RevModPhys.69.315}.

\end{thebibliography}

\end{document}